\documentclass[preprint,floats,aps,12pt]{revtex4}
\usepackage{times} 
\usepackage{amssymb}
\usepackage{amsmath}
\usepackage{prelim2e}\usepackage[none,bottom]{draftcopy}
\draftcopyName{UT preprint / }{1.2}

\ifx\pdfoutput\undefined
\usepackage[usenames,dvips]{color}
\usepackage{graphicx}     
\else
\usepackage[usenames,dvipsnames]{color}
\usepackage{epstopdf}
\usepackage[pdftex]{graphicx}
\usepackage[pdftex]{hyperref}
\hypersetup{a4paper,
    pdftitle={Loughborough Manuscrito}
    pdfauthor={Uwe Thiele},
pdfproducer={lateX},
pdfview=FitV,       
pdfstartview=FitB,
linkcolor=blue,     
pagecolor=blue,     
citecolor=red,      
urlcolor=red,      
breaklinks=true,    
colorlinks=true,
citebordercolor=0 0 0,  
filebordercolor=0 0 0,
linkbordercolor=0 0 0,
menubordercolor=0 0 0,
pagebordercolor=0 0 0,
urlbordercolor=0 0 0,
pdfhighlight=/I,
pdfborder=0 0 0,   
backref=false,
pagebackref=false,
bookmarks=true,
bookmarksopen=true,
bookmarksnumbered=true
}
\fi
\ifx\pdfoutput\undefined
\DeclareGraphicsExtensions{.eps}
\else
\DeclareGraphicsExtensions{.jpg, .pdf, .png, .tif}
\fi
\usepackage{color}
\graphicspath{{./Figures/}} 
\newcommand{\mylab}[1]{\label{#1}  }     
\usepackage[usenames,dvipsnames]{color}

\begin{document}

\title{Localized states in the conserved Swift-Hohenberg
  equation with cubic nonlinearity}
\author{Uwe Thiele}
\email{u.thiele@lboro.ac.uk}
\homepage{http://www.uwethiele.de}
\author{Andrew J. Archer}
\author{Mark J. Robbins}
\affiliation{Department of Mathematical Sciences, Loughborough University,
Loughborough, Leicestershire, LE11 3TU, UK}
\author{Hector Gomez}
\affiliation{University of A Coruna, Campus de Elvina, 15192 A Coruna, Spain}
\author{Edgar Knobloch}
\affiliation{Department of Physics, University of California, Berkeley, California 94720, USA}
\begin{abstract}
The conserved Swift-Hohenberg equation with cubic nonlinearity provides the
simplest microscopic description of the thermodynamic transition from a fluid state 
to a crystalline state. The resulting phase field crystal model describes a 
variety of spatially localized structures, in addition to different spatially extended 
periodic structures. The location of these structures in the temperature versus mean order
parameter plane is determined using a combination of numerical continuation in one
dimension and direct numerical simulation in two and three dimensions. Localized
states are found in the region of thermodynamic coexistence between the homogeneous
and structured phases, and may lie outside of the binodal for these states. The
results are related to the phenomenon of slanted snaking but take the form of standard 
homoclinic snaking when the mean order parameter is plotted as a function of the chemical
potential, and are expected to carry over to related models with a conserved order parameter.
\end{abstract}
%
%
\maketitle
%
%
\section{Introduction} \mylab{intro}
%
Spatially localized structures (hereafter LS) are observed in a great
variety of pattern-forming systems. Such states include spot-like 
structures found in reaction-diffusion systems \cite{CRT00} and in a 
liquid light-valve experiment \cite{BCR09}, isolated
spikes observed in a ferrofluid experiment \cite{RB05} and localized buckled 
states resulting from the buckling of slender structures under compression 
\cite{HPCWWBL00}. Related states are observed in fluid mechanics, including 
convection \cite{bkam06,BK:08b,Blanchflower99} and shear flows \cite{sgb10}.
In other systems, such as Faraday waves, the LS oscillate in time,
either periodically or with a more complex time-dependence, forming
structures referred to as oscillons \cite{laf96,rlc11}. This is also
the case for oscillons in granular media \cite{UMS:96} and in
optics \cite{AFO09}. Other examples of LS include localized traveling
waves \cite{Kolodner88,Barten91} and states called ``worms'' observed in
electroconvection \cite{Dennin96,Riecke98}. In many of these systems the
use of envelope equations removes the (fast) time-dependence and maps
such time-dependent structures onto equilibria. 

Many of the structures mentioned above are examples of ``dissipative
solitons'' \cite{PBA10}
in which energy loss through dissipation is balanced by energy input through 
spatially homogeneous forcing. Others (e.g.,~\cite{RB05,HPCWWBL00}) correspond 
to local minima of an underlying energy or Lyapunov functional. This is the
case for systems whose dynamics are of gradient type, provided only that the 
energy is bounded from below. On finite domains with null boundary conditions 
such systems evolve to steady states which may or may not 
correspond to LS. However, in generic systems supporting dissipative solitons
no Lyapunov function will be present and the evolution of the system may 
be nonmonotonic.
 
In either system type, time-independent LS are found in regions of parameter 
space in which a time-independent spatially homogeneous (i.e., uniform) state 
coexists with a time-independent spatially periodic state. Within this region 
there is a subregion referred to as the {\it snaking} or {\it pinning} region
\cite{Pom:86} in which a great variety of stationary LS are present. In
the simplest, one-dimensional, case the LS consist of a segment of the
periodic state embedded in a homogeneous background; segments of the
homogeneous state embedded in periodic background may be thought of as
localized hole states LH. Steady states of this type lie on a pair of intertwined branches 
that snake back and forth across the snaking region, one of which consists of 
reflection-symmetric LS with a peak in the center, while the other consists 
of similar states but with a dip in the center, and likewise for the holes. 
Near the left edge of the snaking region each LS adds a pair of new cells, 
one at either end, and these grow to full strength
as one follows the LS across the snaking region to its right edge,
where both branches turn around and the process repeats. Thus, as one
follows the LS branches towards larger $L^2$ norm, both types of LS
gradually grow in length, and all such structures coexist within the
snaking region. On a finite interval the long LS take the form of holes
in an otherwise spatially periodic state but on the real line, the LS and 
LH remain distinct although both occupy the same snaking region.
The LS branches are, in adition, interconnected 
by cross-links resembling the rungs of a ladder, consisting
of asymmetric LS \cite{bukn07}.  In generic systems posed on the real
line states of this type drift, either to the left or the right,
depending on the asymmetry, but in systems with gradient dynamics the
asymmetric states are also time-independent.  These, along with bound
states of two, three, etc LS/LH are also present within the snaking
region \cite{bukn07}.

The above behavior is typical of nonconserved order parameter fields.
However, an important subclass of gradient systems possesses a conserved 
quantity, and in such systems the order parameter field has a fixed mean
value. Systems of this type arise frequently in fluid convection and 
other applications \cite{CM01,LBK11,BBKK12} and are distinguished from 
the standard scenario summarised above by the following properties 
\cite{FCS07,dawe08}: (i) the snaking becomes slanted (sometimes referred 
to as ``sidewinding''), (ii) LS may be present outside of the region of 
coexistence of the homogeneous and periodic states, (iii) LS are present 
even when the periodic states bifurcate supercritically, i.e., when the 
coexistence region is absent entirely. The slanting of the snakes-and-ladders 
structure is a finite size effect: in a finite domain expulsion of the 
conserved quantity from the LS implies its shortage outside, a fact 
that progressively delays (to stronger forcing) the formation
events whereby the LS grow in length. The net effect is that LS are found 
in a much broader region of parameter space than in nonconserved systems.

The above properties are shared by many of the models arising in dynamical
density functional theory (DDFT) and related phase field models of crystalline 
solids. The simplest such phase field crystal (PFC) model \cite{PFC_review} 
(see below) leads to the so-called conserved Swift-Hohenberg (cSH) equation. 
This equation was first derived, to the authors' knowledge, as the equation 
governing the evolution of binary fluid convection between thermally insulating
boundary conditions \cite{K:89}; for recent derivations in the PFC context see 
Refs.~\cite{vBVL09, ARTK12,PFC_review}. In this connection the PFC model may be
viewed as probably the simplest {\em microscopic} model for the freezing
transition that can be constructed. In this model the transition from a 
homogeneous state to a periodic state corresponds to the transition from a 
uniform density liquid to a periodic crystalline solid. The LS of interest in 
this model then correspond to states in which a finite size portion of the 
periodic crystalline phase coexists with the uniform density liquid phase, and 
these are expected to be present in the coexistence region between the two 
phases. Some rather striking examples of LS in large two-dimensional systems 
include snow-flake-like and dendritic structures, e.g., Refs.\ 
\cite{PFC_review, tegz09,tegz09b,tegz09c,tams08}. In fact, as shown below, the 
LS are also present at state points outside of the coexistence region. However,
despite the application of the cSH (or PFC) equation in this and other areas, 
the detailed 
properties of the LS described by this equation have not been investigated. 
In this paper we make a detailed study of the properties of this equation in 
one spatial dimension with the aim of setting this equation firmly in the body 
of literature dealing with spatially localized structures. Our results are 
therefore interpreted within both languages, in an attempt to make existing 
understanding of LS accessible to those working on nonequilibrium models of 
solids, and to use the simplest PFC model to exemplify the theory. In addition,
motivated by Refs.\ \cite{tegz09,tegz09b,tegz09c,tams08}, we also describe 
related results in two (2d) and three (3d) dimensions, where (many) more types 
of regular patterns are present and hence many more types of LS. Our work
focuses on `bump' states (also referred to as `spots') which are readily
found in direct numerical simulations of the conserved Swift-Hohenberg
equation as well as in other systems \cite{BCR09}.

Although the theory for these cases in 2d and 3d is less well
developed \cite{LSAC08,ALBKS10} continuation results indicate that
some of the various different types of LS can have quite different
properties. For example, the bump states
differ from the target-like LS formed from the stripe state
that can also be seen in the model. In particular, spots in the
nonconserved Swift-Hohenberg equation in the plane bifurcate from the
homogeneous state regardless of whether stripes are subcritical or
supercritical \cite{LS09}, see also Ref.~\cite{MS10}.  The key
question, hitherto unanswered, is whether two-dimensional structures
in the plane snake indefinitely and likewise for three-dimensional
structures.

The paper is organized as follows. 
In Sec. II we describe the conserved Swift-Hohenberg equation and its basic 
properties. In Sec. III we describe the properties of LS in one spatial 
dimension as determined by numerical continuation. In Sec. IV we describe 
related results in two and three spatial dimensions, but obtained by direct 
numerical simulation of the PFC model. Since  this model has a gradient 
structure, on a finite domain all solutions necessarily approach a 
time-independent equilibrium. However, as we shall see, the number of 
competing equilibria may be very large and different equilibria are reached 
depending on the initial conditions employed. In Sec. V we put our results 
into context and present brief conclusions.

\section{The conserved Swift-Hohenberg equation}
\subsection{Equation and its variants}
\mylab{sec:eqs}

We write the cSH (or PFC) equation in the form
 \begin{equation}
\partial_t \phi(\mathbf{x},t)=\alpha \nabla^2 \frac{\delta F[\phi]}{\delta \phi(\mathbf{x},t)} ,
\mylab{eq:DDFT_PFC}
\end{equation}
where $\phi(\mathbf{x},t)$ is an order parameter field that corresponds in the PFC context to a scaled density profile, $\alpha$ is a (constant) mobility coefficient and $F[\phi]$ denotes the free energy functional
\begin{eqnarray}
F[\phi]\equiv \int d\mathbf{x} \left[\frac{\phi}{2}[r+(q^2+\nabla^2)^2]\phi+\frac{\phi^4}{4}\right].
\mylab{eq:hfe}
\end{eqnarray}
Here $\mathbf{x}=(x,y,z)$, $\nabla=(\partial_x,\partial_y,\partial_z)^T$ is 
the gradient operator, and subscripts denote partial derivatives. It follows 
that the system evolves according to the cSH equation
\begin{equation}
  \partial_t\,\phi\,=\, \alpha \nabla^2\left[r\phi + (\nabla^2+q^2)^2\phi + \phi^3\right].
\mylab{eq:csh}
\end{equation}

Equation (\ref{eq:csh}) is sometimes called the derivative Swift-Hohenberg
equation \cite{MaCo00,Cox04}; many papers use a different sign convention 
for the parameter $r$ (e.g.,
\cite{EKHG02,Achi09,GDL09,SHP06,BRV07,OhSh08,MKP08,tegz09,vBVL09}).
In this equation the quartic term in $F[\phi]$ may be replaced by other types
of nonlinearity, such as $f_{23}=-b_2\phi^3/3+\phi^4/4$ 
\cite{MaCo00,Cox04,tegz09,vBVL09} without substantial change in
behavior. Related but nonconserved equations $\partial_t
\phi=-\tilde\alpha\delta F[\phi]/\delta \phi$ with nonlinear
terms of the form $f_{23}$ \cite{BuKn06} or $f_{35}=-b_3\phi^4/4+\phi^6/6$
\cite{bukn07} have also been extensively studied, subject to the
conditions $b_2>27/38$ (resp., $b_3>0$) required to guarantee
the presence of an interval of coexistence between the homogeneous
state $\phi=0$ and a spatially periodic state. Note that in the
context of nonconserved dynamics \cite{BuKn06,bukn07} one generally
selects a nonlinear term $g_\mathrm{nl}$ directly, although this term
is related to $f_\mathrm{nl}$ through the relation
$g_\mathrm{nl}\equiv -d f_\mathrm{nl}/d\phi$, i.e., $g_{23}$ or $g_{35}$.
As we shall see below, in the conserved case, having the nonlinear term 
$f_{23}$ describes the generic case and the role of the coefficient $b_2$ 
is effectively played by the value of $\phi_0$, which is the average 
value of the order parameter $\phi(\mathbf{x})$.

Equation (\ref{eq:csh}) can be studied in one, two or more dimensions. In 
one dimension with $g_{23}$ the equation was studied by Matthews and Cox 
\cite{MaCo00,Cox04} as an example of a system with a conserved order parameter;
this equation is equivalent to Eq.~(\ref{eq:csh}) with imposed nonzero mean 
$\phi$. A weakly localized state of the type that is of interest in the 
present paper is computed in \cite{MaCo00} and discussed further in 
\cite{Cox04}.

\subsection{Localized states in one spatial dimension}
\mylab{sec:loc-states-1d}

We first consider Eq.~(\ref{eq:csh}) in one dimension, with $\alpha=1$ and $q=1$, i.e.,
\begin{equation}
  \partial_t\,\phi\,=\, \partial^2_x\left[r\phi +
      (\partial^2_{x}+1)^2\phi +\phi^3\right].
\mylab{eq:csh-loc}
\end{equation}
This equation is reversible in space (i.e., it is invariant under 
$x\rightarrow -x$). Moreover, it conserves the total ``mass'' 
$\int_0^L\phi\,dx$, where $L$ is the size of the system. In the following 
we denote the average value of $\phi$ in the system by 
$\phi_0\equiv\langle\phi\rangle$ so that perturbations 
${\tilde\phi}\equiv \phi-\phi_0$
necessarily satisfy $\langle{\tilde\phi}\rangle=0$, where 
$\langle\cdots\rangle\equiv L^{-1}\int_0^L(\cdots)\,dx$.

Steady states ($\partial_t\,\phi\,=\,0$) are solutions of the fourth order
ordinary differential equation
\begin{equation}
 0= r\phi + (\partial_{xx}+1)^2\phi +\phi^3-\mu,
\mylab{eq:csh-loc-steady}
\end{equation}
where $\mu\equiv \delta F[\phi]/\delta\phi$ is an integration constant that 
corresponds to the chemical potential. 

Each solution of this equation corresponds to a stationary value of the underlying Helmholtz free energy 
\begin{equation}
\tilde{F}=\int_0^L\left[(1+r)\frac{\phi^2}{2}+\frac{\phi^4}{4} -
  (\partial_{x}\phi)^2+\frac{1}{2}(\partial_{xx}\phi)^2
\right]\,dx.
\mylab{eq:csh-energy}
\end{equation}
We use the free energy to define the grand potential 
\begin{equation}
\Omega=\tilde{F} - \int_0^L \mu\phi\,dx
\mylab{eq:csh-grand}
\end{equation}
and will be interested in the normalized free energy
density $f=(\tilde{F}[\phi(x)]-\tilde{F}[\phi_0])/L$ and in the
density of the grand potential $\omega=\Omega/L=\tilde{F}[\phi(x)]/L -
\mu{\phi_0}$. We also use the $L^2$ norm
\begin{equation}
||\delta{\phi}||=\sqrt{\frac{1}{L}\int_0^L(\phi-\phi_0)^2\,dx}
\mylab{eq:csh-norm}
\end{equation}
as a convenient measure of the amplitude of the departure of the solution
from the homogeneous background state $\phi=\phi_0$.

Linearizing Eq.~(\ref{eq:csh-loc}) about the
steady homogeneous solution $\phi=\phi_0$ using the ansatz
$\delta\phi(x,t)\equiv\phi(x,t)-\phi_0=\epsilon\exp(\beta t + i kx)$ with $\epsilon\ll1$ results in 
the dispersion relation
\begin{equation}
\beta=-k^2\,[r+(1-k^2)^2+3\phi_0^2].
\mylab{eq:csh-disp}
\end{equation}
It follows that in an infinite domain, the threshold for instability of the
homogeneous state corresponds to $r_c^\infty=-3\phi_0^2$. In a domain of finite 
length $L$ with periodic boundary conditions (PBC),
the homogeneous state is linearly unstable for $r<r_n$, where
\begin{equation}
r_n=-(1-k_n^2)^2-3\phi_0^2,
\mylab{eq:csh-stab}
\end{equation}
and $k_n\equiv 2\pi n/L$, $n=1,2,\dots$. 
Standard bifurcation theory with PBC shows that for $L<\infty$ each $r_n$ 
corresponds to a bifurcation point creating a branch of periodic solutions 
that is uniquely specified by the corresponding integer $n$. For those 
integers $n$ for which $r_n > -9/2$ the branch of periodic states bifurcates 
supercritically (i.e., towards smaller values of $|\phi_0|$); for 
$r_n < -9/2$, the bifurcation is subcritical (i.e., the branch bifurcates 
towards larger values of $|\phi_0|$). Since each solution can be translated 
by an arbitrary amount $d$ 
(mod $L$), each bifurcation is in fact a pitchfork of revolution. Although 
the periodic states can be computed analytically for $r\approx r_n$, for
larger values of $|r-r_n|$ numerical computations are necessary. In the 
following we use the continuation toolbox AUTO \cite{DKK91,AUTO07P} 
to perform these (and other) calculations. For interpreting the results it
is helpful to think of $r$ as a temperature-like variable which
measures the undercooling of the liquid phase.

\begin{figure}
\includegraphics[width=0.8\hsize]{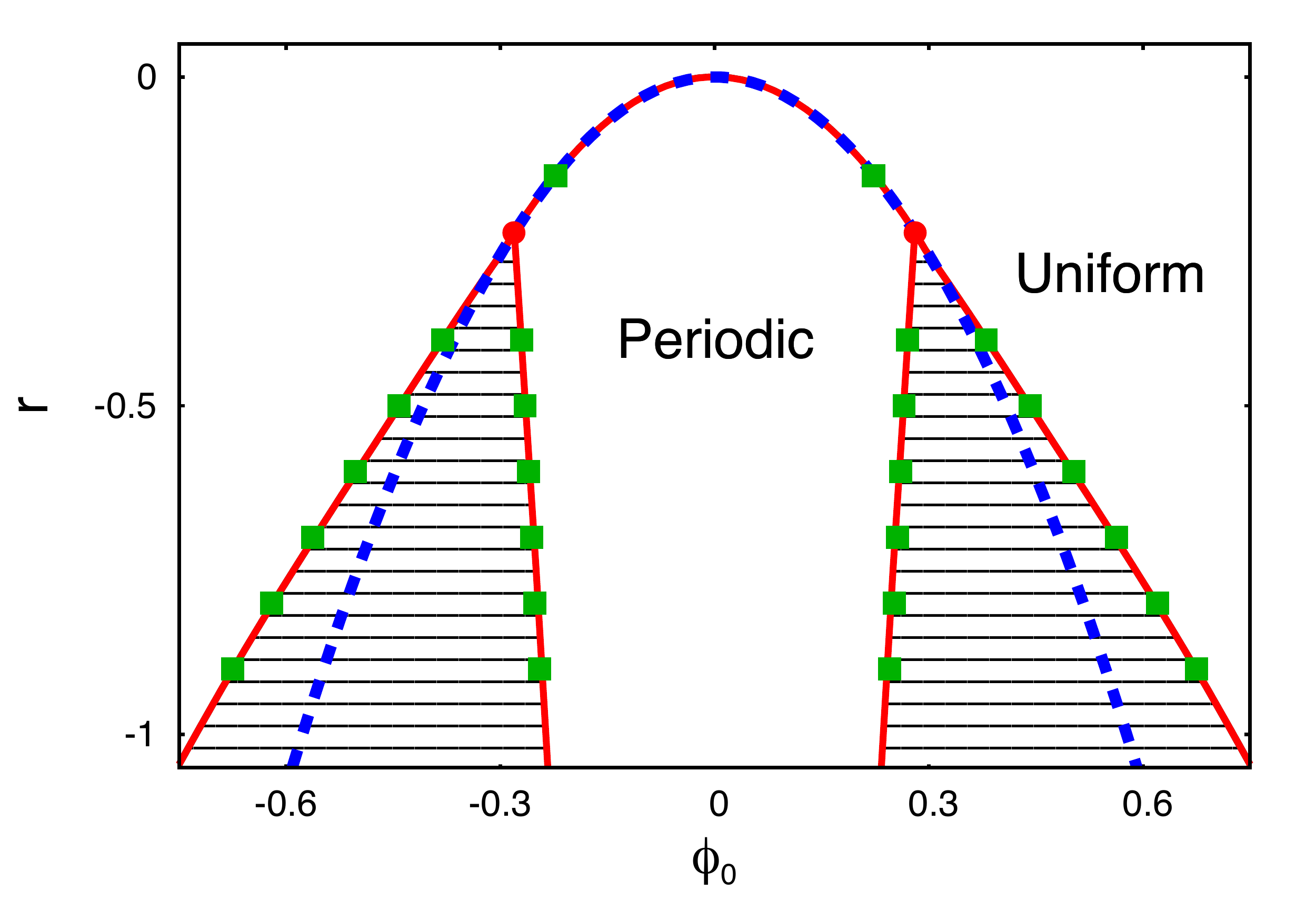}
\caption{(color online) The phase diagram for the 1d PFC model (\ref{eq:csh})
  when $q = 1$. The red solid lines are the coexistence
  curves between the periodic and uniform phases calculated using a
  two mode approximation \cite{ARTK12}. The green squares show the coexistence
  values calculated from simulations \cite{ARTK12}. The red circles are the
  tricritical points. The blue dashed line shows the curve of marginal 
  stability of the uniform state within linear stability theory.}
\mylab{fig:cSH-phasediagram-1d}
\end{figure}

Before we can discuss LS in the above model, it is helpful to refer to the 
phase diagram appropriate to a one-dimensional setting 
(Fig.~\ref{fig:cSH-phasediagram-1d}).   As shown in
Ref.~\cite{ARTK12} the tricritical point is located at
$(\phi_{0b},r_b^\mathrm{max})=(\pm\sqrt{3/38},-9/38)$. For
$r>r_b^\mathrm{max}$ there exists no thermodynamic coexistence zone
between the homogeneous and periodic states. Such a region
is only present for $r<r_b^\mathrm{max}$ and is limited by the binodal
lines that indicate the values of $\phi_0$ for which the homogeneous
and periodic solutions at fixed $r$ have equal chemical potential and
pressure (i.e., equal grand potential). Thus for $r<r_b^\mathrm{max}$
the transition from the homogeneous to the periodic state is of first 
order. The binodals can
either be calculated for specific domain sizes and periods of the
periodic structure or for an infinite domain. In the latter case the
period of the periodic state is not fixed but corresponds to the
period that minimizes the Helmholtz free energy at each $(\phi_0,r)$
\cite{ARTK12}. We remark that with this choice of parameters the 
tricritical point is not the point at which the bifurcation to the 
periodic state changes from supercritical to subcritical. As already 
mentioned, the latter occurs at $(\phi_0,r)=(\pm\sqrt{3/2},-9/2)$, 
i.e., at values of $r$ much smaller than $r_b^\mathrm{max}$. Further 
discussion of this point may be found in the conclusions.

\section{Results for the conserved Swift-Hohenberg equation}
\subsection{Families of localized states}

Since Eq.~(\ref{eq:csh-loc}) represents conserved gradient dynamics based on
an energy functional that allows for a first order phase transition
between the homogeneous state and a periodic patterned state, one may
expect the existence of localized states (LS) to be the norm rather than
an exception. In the region between the binodals, where homogeneous and periodic
structures may coexist, the value of $\phi_0$, i.e., the amount of `mass' in the
system, determines how many peaks can form.

As in other problems of this type we divide the LS into three classes. 
The first class consists of left-right symmetric structures with a peak
in the middle. Structures of this type have an overall odd number of 
peaks and we shall refer to them as odd states, hereafter LS$_\mathrm{odd}$. 
The second class consists of left-right symmetric structures with a dip 
in the middle. Structures of this type have an overall even number of 
peaks and we refer to them as even states, hereafter LS$_\mathrm{even}$. 
Both types have even parity with respect to reflection in the center 
of the structure. The third class consists of states of no fixed parity, 
i.e., asymmetric states, LS$_\mathrm{asym}$. The asymmetric states are
created from the symmetric states at pitchfork bifurcations and take the 
form of rungs on a ladder-like structure with
interconnections between LS$_\mathrm{odd}$ to 
LS$_\mathrm{even}$. In view of the gradient structure of Eq.~(\ref{eq:csh-loc}) 
the asymmetric states are likewise stationary solutions of the equation.

We now address the following questions:\\
1. Do localized states exist outside the binodal region? Can they form
the energetic minimum outside the binodal region?\\
2. How does the bifurcational structure of the localized states change
with changes in the temperature-like parameter $r$? How does the transition from 
tilted or slanted snaking to no snaking occur? What is the behavior of the
asymmetric localized states during this process?

Answers to these and other questions can be obtained by means of an
in-depth parametric study. In the figures that follow we present
bifurcation diagrams for localized states as a function of the mean
order parameter value $\phi_0$ for a number of values of the
parameter $r$. All are solutions of Eq.\ \eqref{eq:csh-loc-steady} that
satisfy periodic boundary conditions on
the domain $0\le x\le L$, and are characterized by their $L^2$ norm
$||\delta \phi||$, chemical potential $\mu$, free energy density $f$,
and grand potential density $\omega$ as defined in
Sec.~\ref{sec:loc-states-1d}.

In Fig.~\ref{fig:loc-fam-rm09} we show the results for $r=-0.9$ for $L=100$. 
Figure \ref{fig:loc-fam-rm09}(a) shows $||\delta\phi ||$ as a function of 
$\phi_0$: the classical bifurcation diagram. For these parameter values, 
as $\phi_0$ is increased the homogeneous (liquid) phase first becomes 
unstable to perturbations with mode number $n=16$ (i.e., 16 bumps), 
followed closely by bifurcations to modes with $n=15$ and $n=17$. 
All other modes bifurcate at yet smaller values of $|\phi_0|$ and are
omitted. All three primary bifurcations are supercritical. The figure
also reveals that the $n=16$ branch undergoes a secondary instability
already at small amplitude; this instability creates a pair of
secondary branches of spatially localized states, LS$_\mathrm{odd}$
(solid line) and LS$_\mathrm{even}$ (dashed line). With increasing
amplitude these branches undergo {\it slanted snaking} as one would expect 
on the basis of the results for related systems with a conservation law
\cite{dawe08,LBK11}. The LS$_\mathrm{odd}$ and LS$_\mathrm{even}$
branches are in turn connected by ladder branches consisting of
asymmetric states LS$_\mathrm{asym}$, much as in standard snaking
\cite{BuKn06}. Sample solution profiles for these three types of LS
are shown in Fig.~\ref{fig:loc-prof-rm09}.  The snaking ceases when
the LS have grown to fill the available domain; in the present case
the LS$_\mathrm{odd}$ and LS$_\mathrm{even}$ branches terminate on
the same $n=16$ branch that created them in the first place.
Whether or not this is the case depends in general on the domain
length $L$, as discussed further in Ref.~\cite{BBKM08}.

The key to the bifurcation diagram shown in Fig.~\ref{fig:loc-fam-rm09}(a)
is evidently the small amplitude bifurcation on the $n=16$ branch. This
bifurcation destabilises the $n=16$ branch that would otherwise be stable 
and is a consequence of the presence of the conserved quantity $\phi_0$
\cite{MaCo00}. As $L$ increases, the bifurcation moves down to smaller and
smaller amplitude, so that in the limit $L\rightarrow\infty$ the periodic branch
is entirely unstable and the LS bifurcate directly from the homogeneous
state. Since the LS bifurcate {\it subcritically} it follows that 
such states are present not only when the primary pattern-forming branch 
is supercritical but moreover are present {\it below} the onset of the 
primary instability.

Figure \ref{fig:loc-fam-rm09}(b) shows the corresponding plot of
the free energy density $f$ as a function of $\phi_0$. This figure 
demonstrates that throughout much of the range of $\phi_0$ the localized 
states have a lower free energy than the extended periodic states. In 
this range the LS are therefore energetically favored. Figure 
\ref{fig:loc-fam-rm09}(c) shows the corresponding plot of the chemical 
potential $\mu$ while Fig.~\ref{fig:loc-fam-rm09}(d) shows the grand 
potential density $\omega$. Of these Fig.~\ref{fig:loc-fam-rm09}(c) is 
perhaps the most interesting since it shows that the results of 
Fig.~\ref{fig:loc-fam-rm09}(a), when replotted using $(\phi_0,\mu)$ to 
characterize the solutions, in fact take the form of standard snaking, 
provided one takes the chemical potential $\mu$ as the control parameter 
and $\phi_0$ as the response. In this form the bifurcation diagram 
gives the values of $\phi_0$ that are consistent with a given value of 
the chemical potential $\mu$ (recall that $\phi_0$ is related to the 
total particle number density). 

\begin{figure}
\textbf{(a)}\includegraphics[width=0.45\hsize]{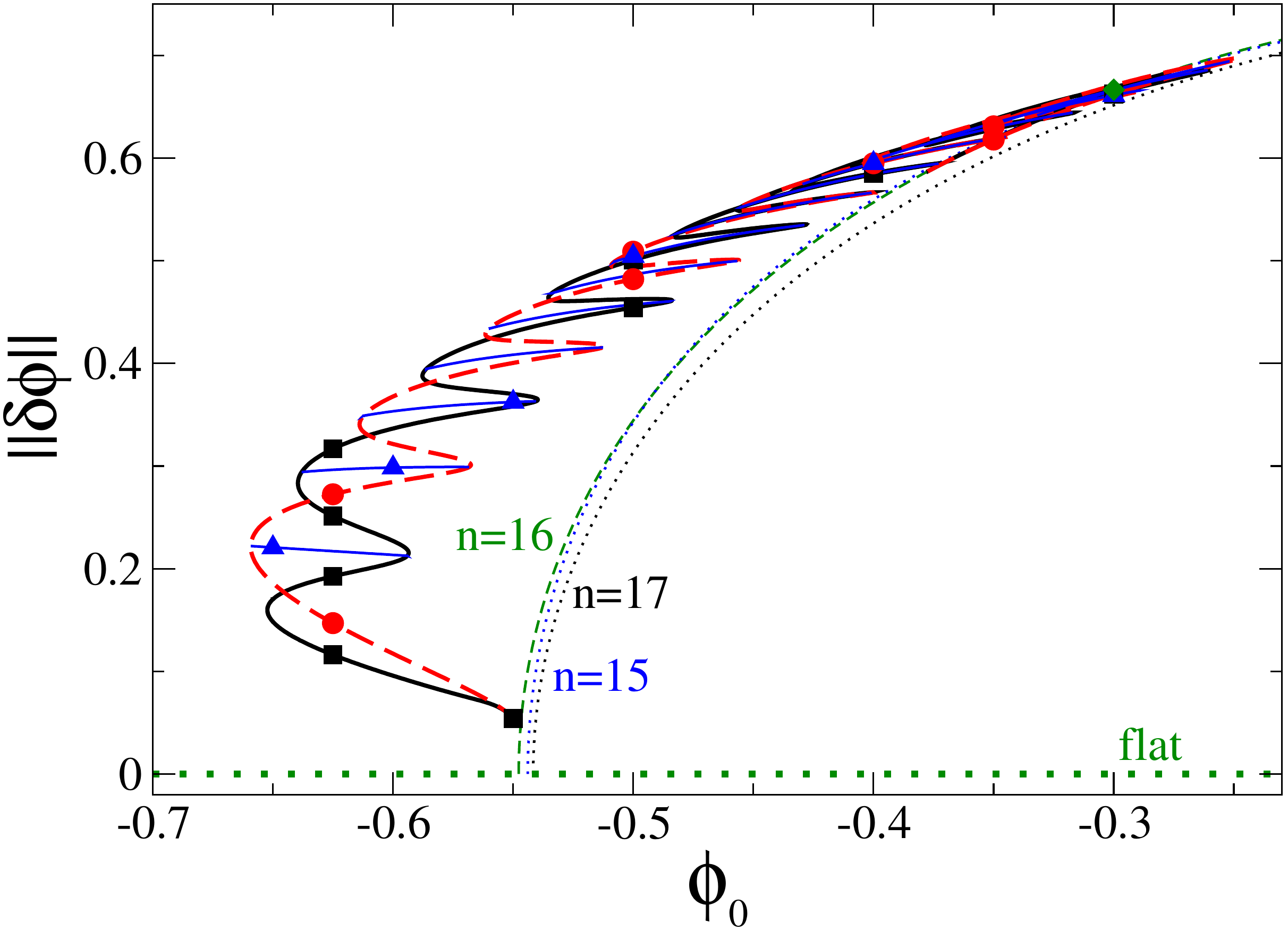}
\includegraphics[width=0.45\hsize]{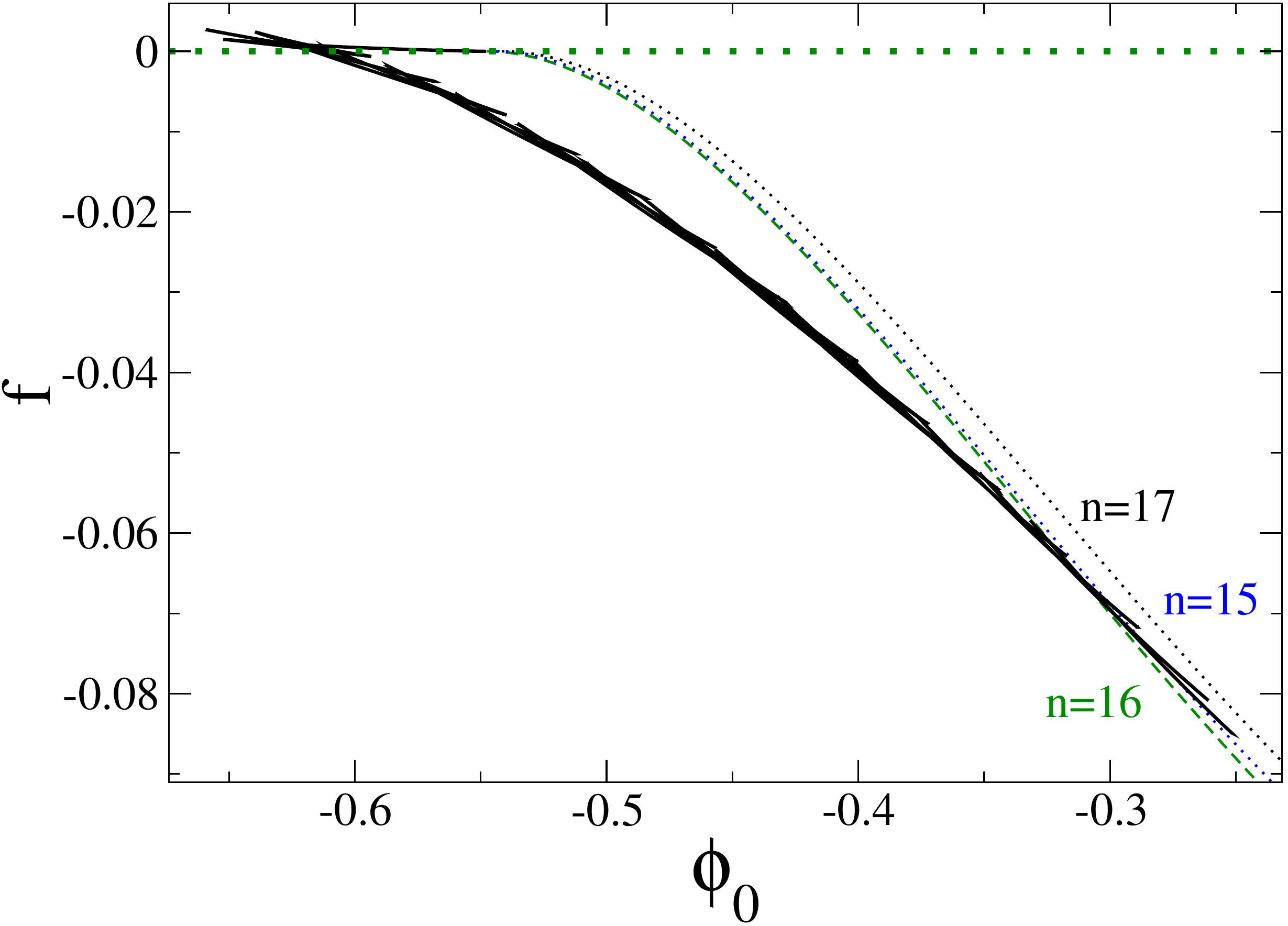}\textbf{(b)}
\textbf{(c)}\includegraphics[width=0.45\hsize]{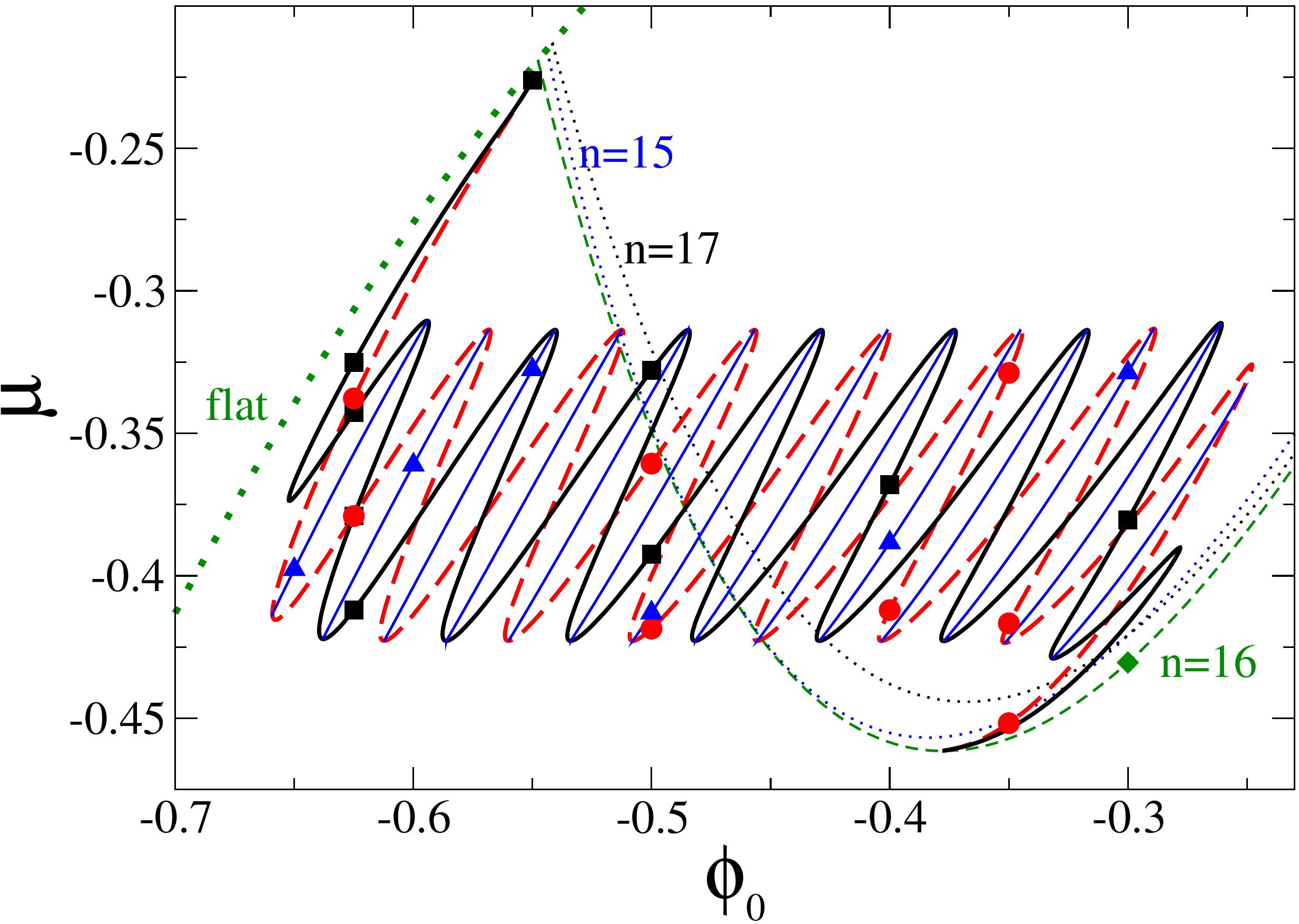}
\includegraphics[width=0.45\hsize]{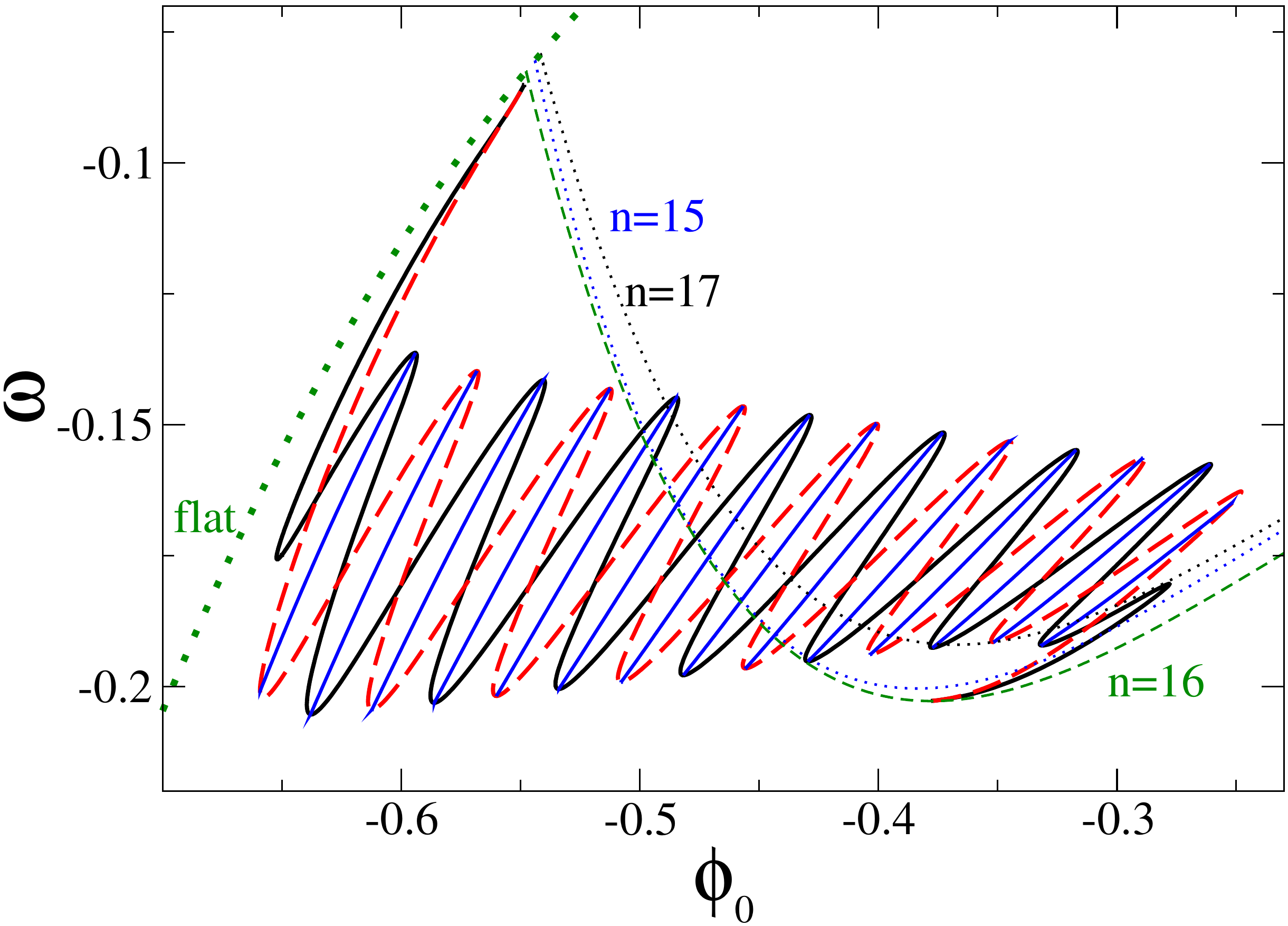}\textbf{(d)}
\caption{(color online) Characteristics of steady state (localized) solutions
  of the one-dimensional conserved
  Swift-Hohenberg equation (\ref{eq:csh-loc}) as a function of the mean
  order parameter $\phi_0$ for a fixed domain size of $L=100$ and
  $r=-0.9$.  The various solution profiles are characterized by their
  (a) $L^2$ norm $||\delta \phi||$, (b) free energy density $f$, (c) chemical 
  potential $\mu$, and (d) grand potential density
  $\omega$.  The thick green dotted line, labeled `flat', corresponds to the
  homogeneous solution $\phi(x)=\phi_0$. Periodic solutions with
  $n=16$ peaks are shown as a thin green dashed line, whereas the nearby
  thin blue and black dotted lines represent the $n=15$ and $n=17$
  solutions, respectively.  The thick solid black and dashed red lines
  that bifurcate from the $n=16$ periodic solution represent symmetric
  localized states with a maximum (LS$_\mathrm{odd}$) and a minimum
  (LS$_\mathrm{even}$) at their center, respectively.  Both terminate on
  the $n=16$ solution. The 14 blue solid lines that connect the
  LS$_\mathrm{odd}$ and LS$_\mathrm{even}$ branches of symmetric
  localized states correspond to asymmetric localized states
  (LS$_\mathrm{asym}$).  Together these three sets of branches of
  localized states form a tilted snakes-and-ladders structure. Typical
  order parameter profiles along the three LS branches are shown in
  Fig.~\ref{fig:loc-prof-rm09}, and correspond to locations indicated
  in panels (a) and (c) by filled black squares (LS$_\mathrm{odd}$),
  red circles (LS$_\mathrm{even}$), blue triangles
  (LS$_\mathrm{asym}$) and green diamond (periodic solution with
  $n=16$).  }
\mylab{fig:loc-fam-rm09}
\end{figure}

\begin{figure}
	\includegraphics[width=0.9\hsize]{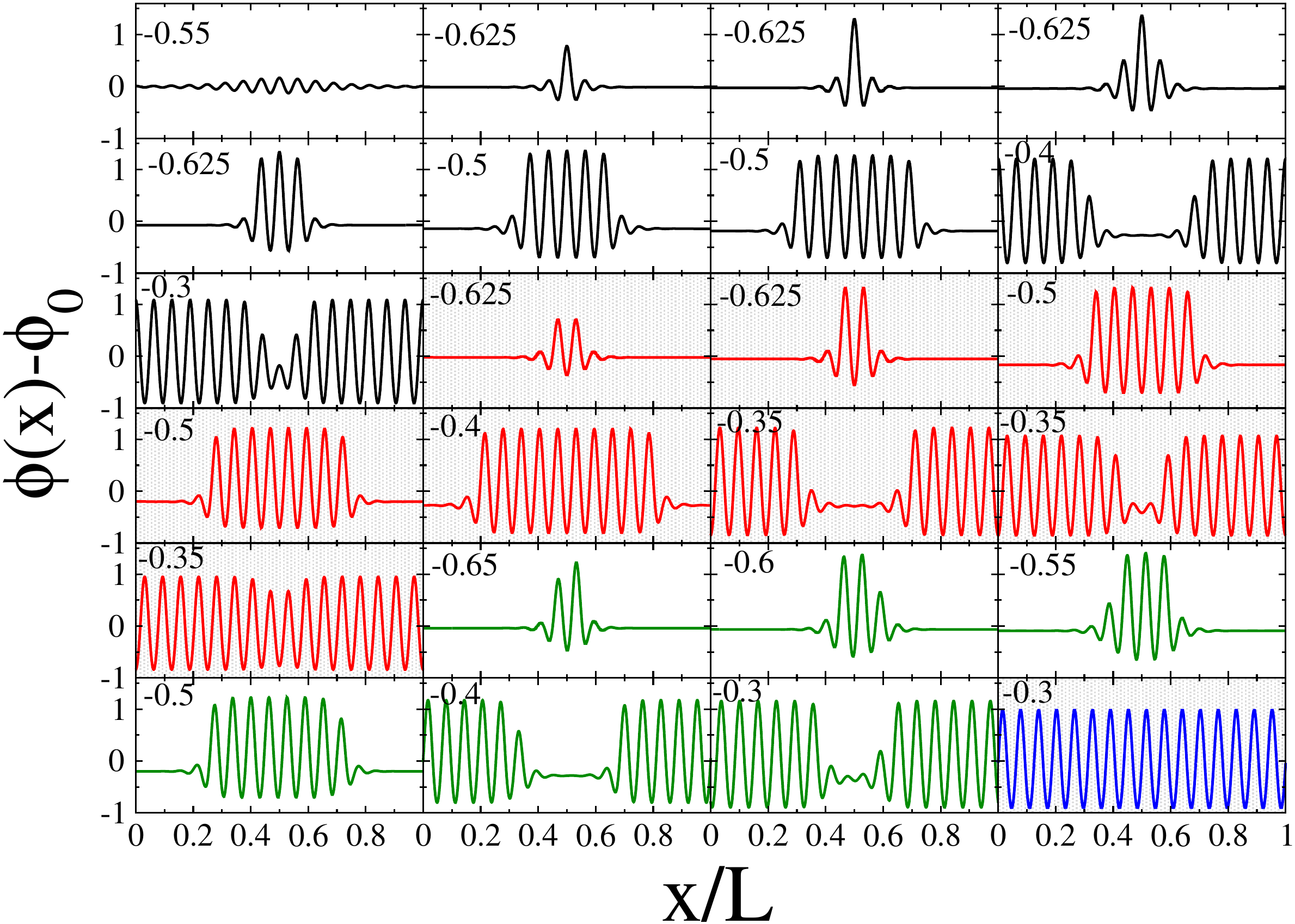}
	\caption{(Color online) A selection of steady state profiles
          $\phi(x)-\phi_0$ for $r=-0.9$ and values of $\phi_0$ in the range
          $-0.65 \leq \phi_0\leq -0.3$ (the number in each panel indicates the corresponding
          value of ${\phi}_0$). Going from top left to bottom right we first
          show nine LS$_\mathrm{odd}$
          solutions, i.e., symmetric localized states with an odd
          number of maxima (in black), then eight LS$_\mathrm{even}$ solutions,
          i.e., symmetric localized states with an even number of
          maxima (in red), followed by six LS$_\mathrm{asym}$ solutions, i.e.,
          asymmetric localized states (in green). The final plot is
          the $n=16$ periodic solution for ${\phi}_0=-0.3$ (in
          blue). The solutions on the symmetric
          branches correspond to locations indicated in 
          Fig.~\ref{fig:loc-fam-rm09}(a) and are shown in order, starting from 
          the bifurcation point that creates them and continuing to larger 
          norm $||\delta \phi||$. The color coding corresponds to that used in 
          Figs.~\ref{fig:loc-fam-rm09}(a,c): LS$_\mathrm{odd}$ (filled black 
          squares), LS$_\mathrm{even}$ (red circles), LS$_\mathrm{asym}$ (blue
          triangles) and periodic (green diamond). }
	\label{fig:loc-prof-rm09}
\end{figure}

\begin{figure}
\vspace*{-1cm}
\textbf{(a)}\includegraphics[width=0.425\hsize]{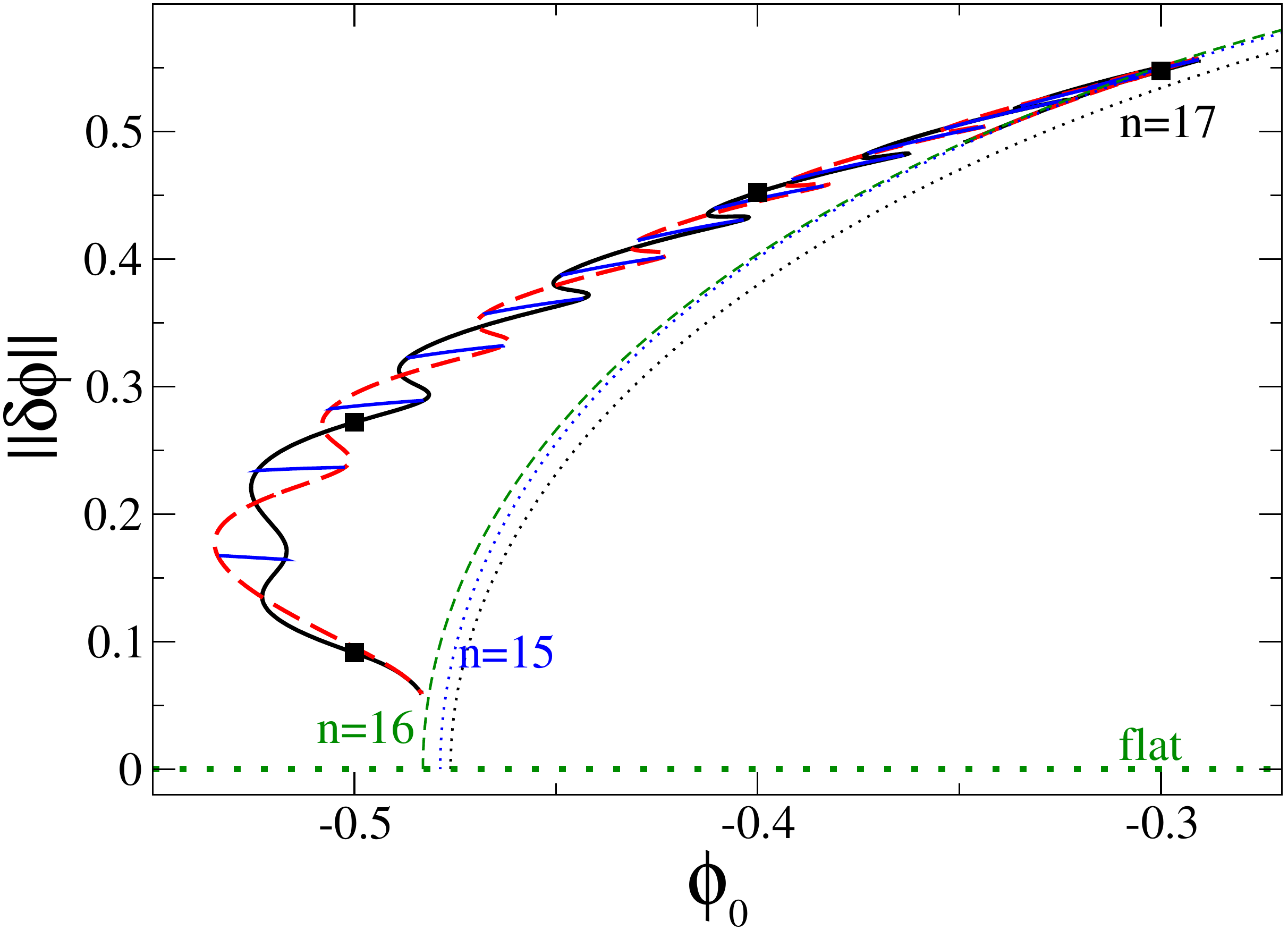}
\includegraphics[width=0.425\hsize]{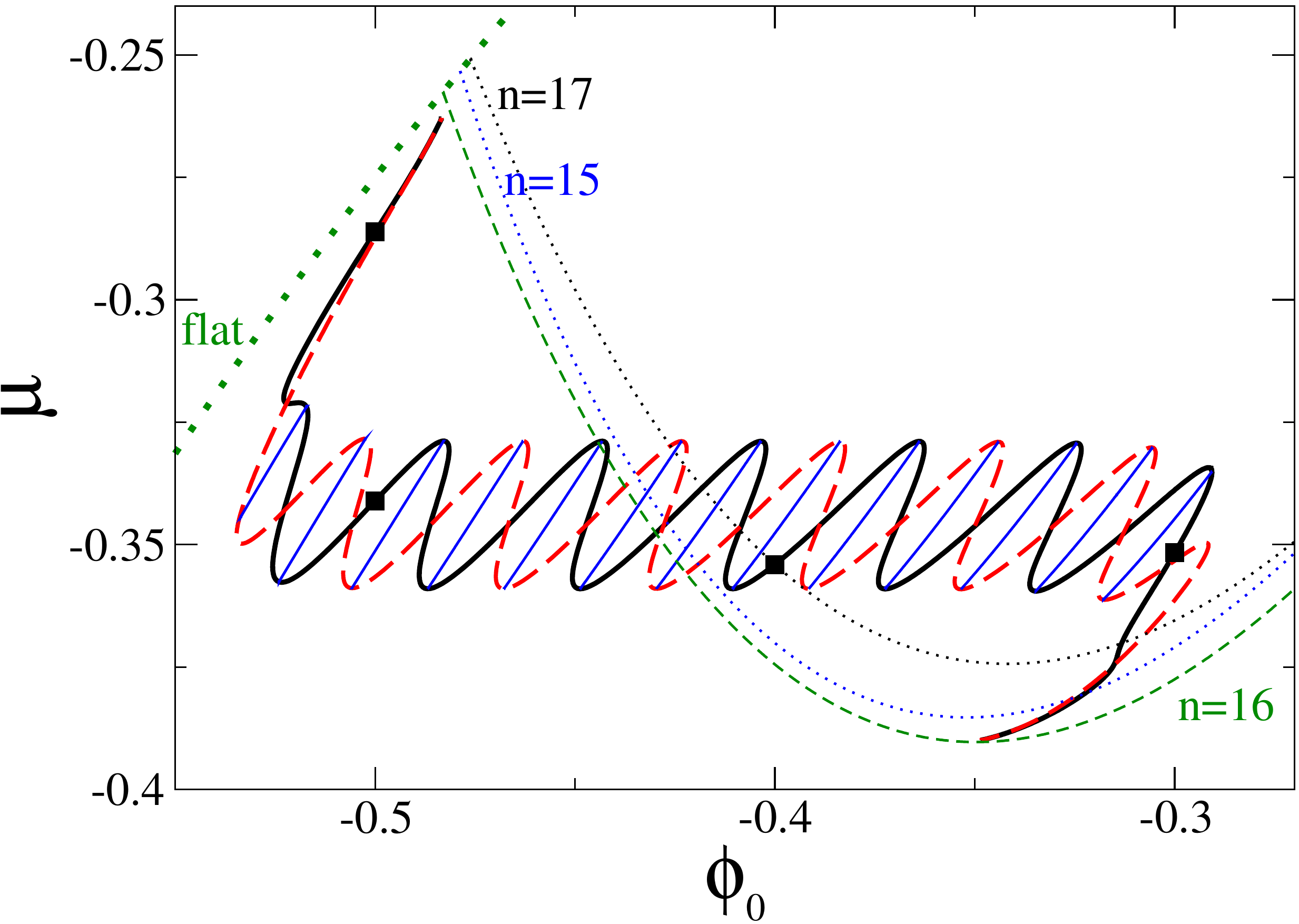}\textbf{(b)}
\textbf{(c)}\includegraphics[width=0.425\hsize]{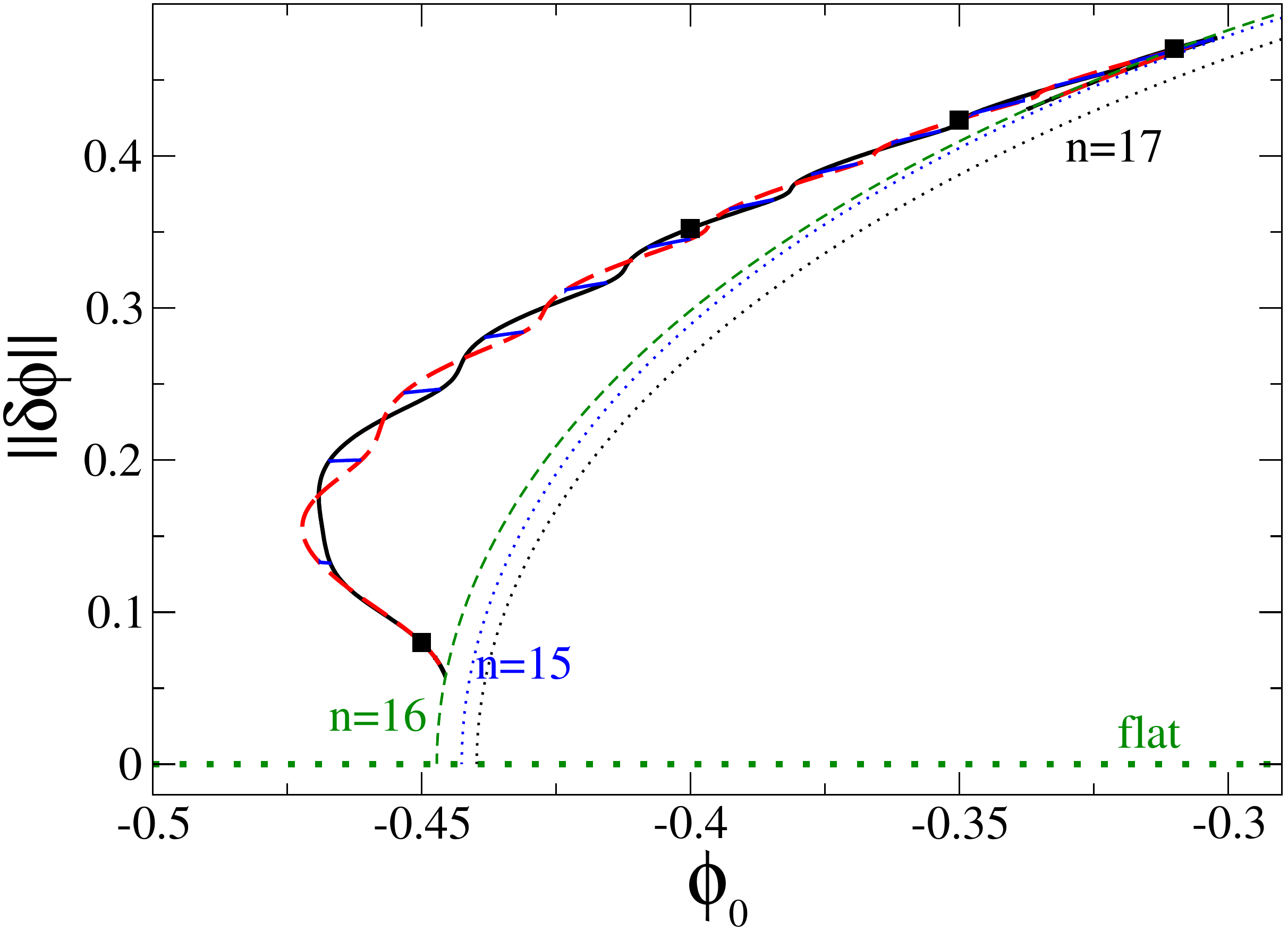}
\includegraphics[width=0.425\hsize]{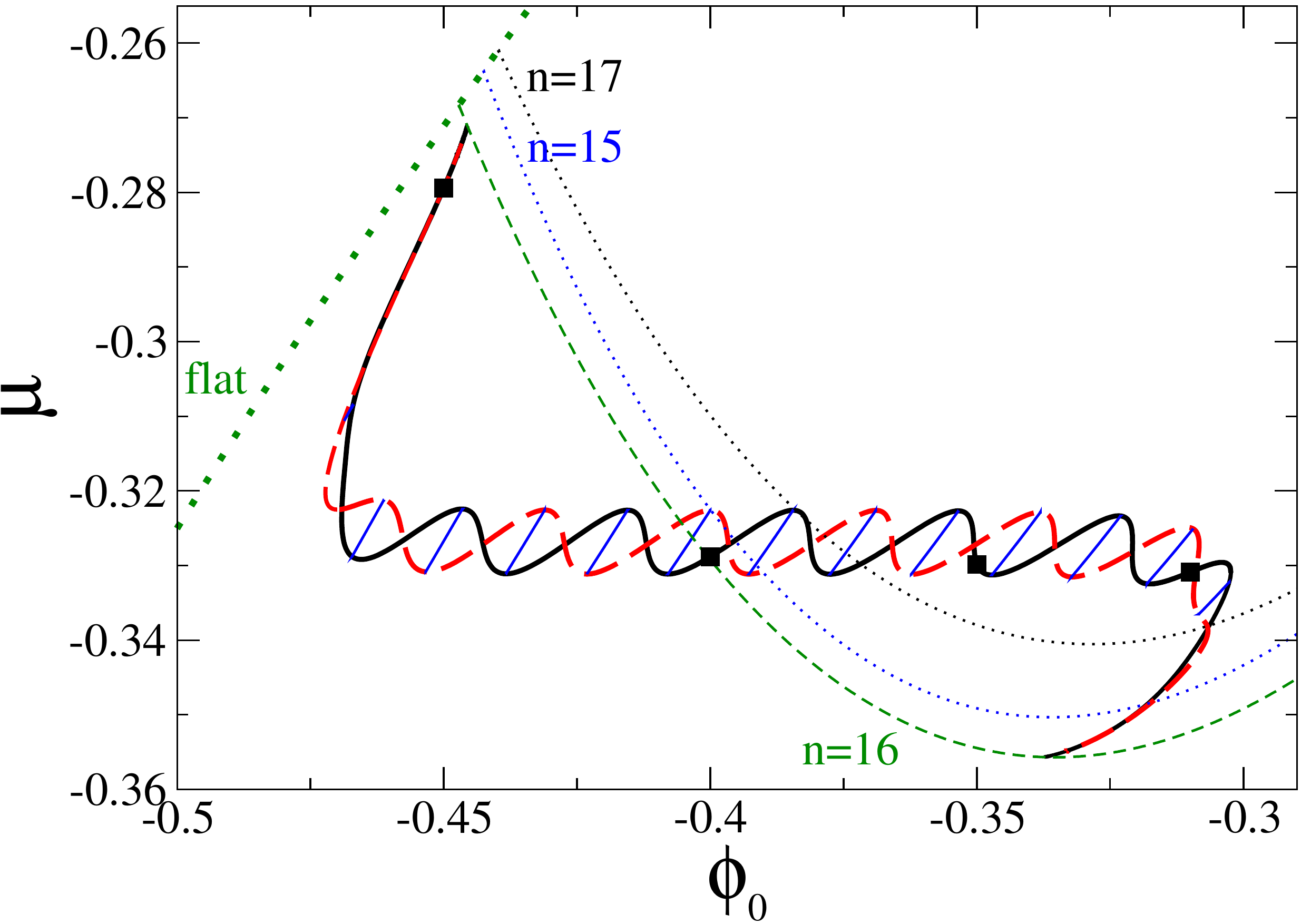}\textbf{(d)}
\textbf{(e)}\includegraphics[width=0.425\hsize]{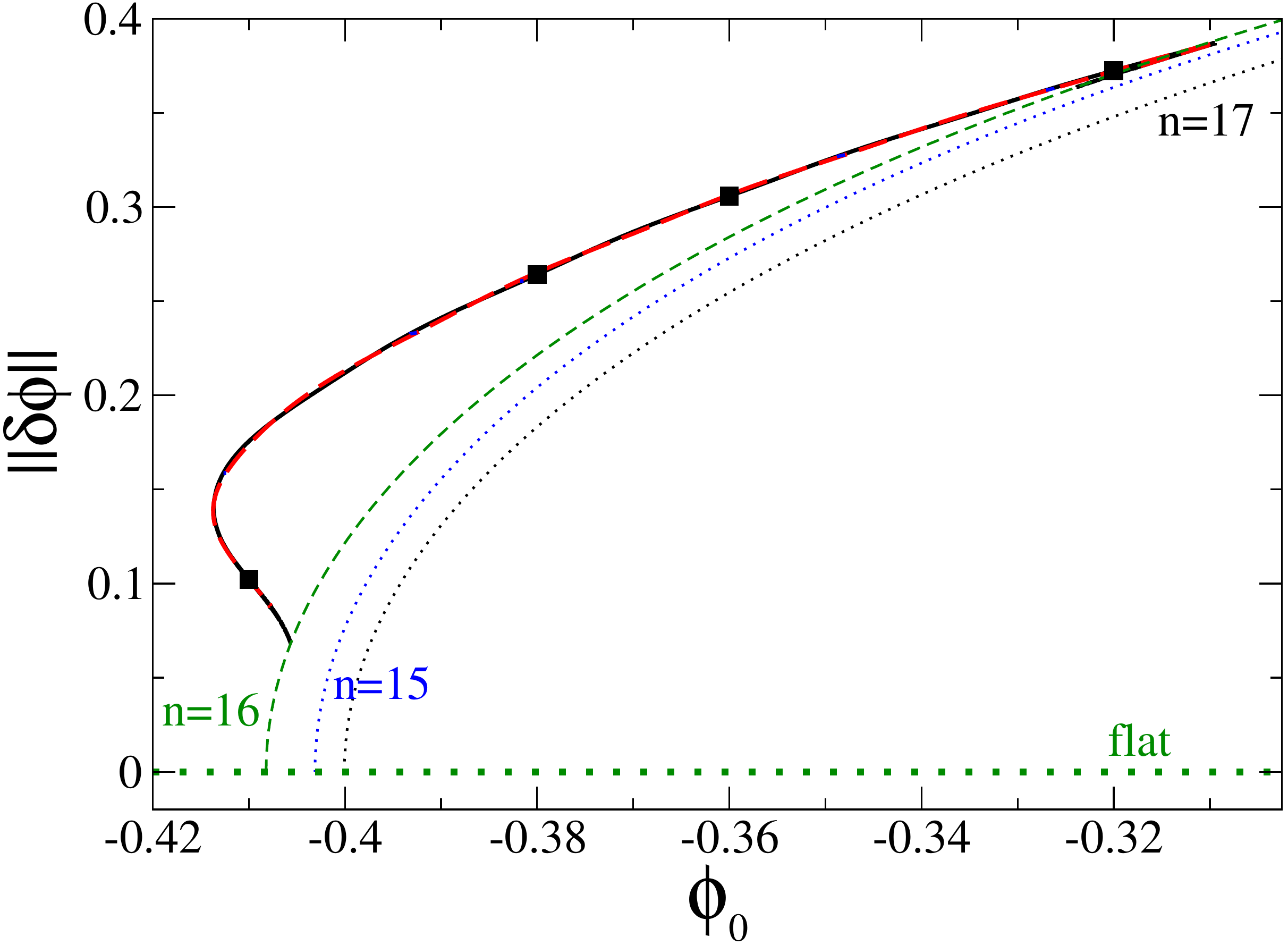}
\includegraphics[width=0.425\hsize]{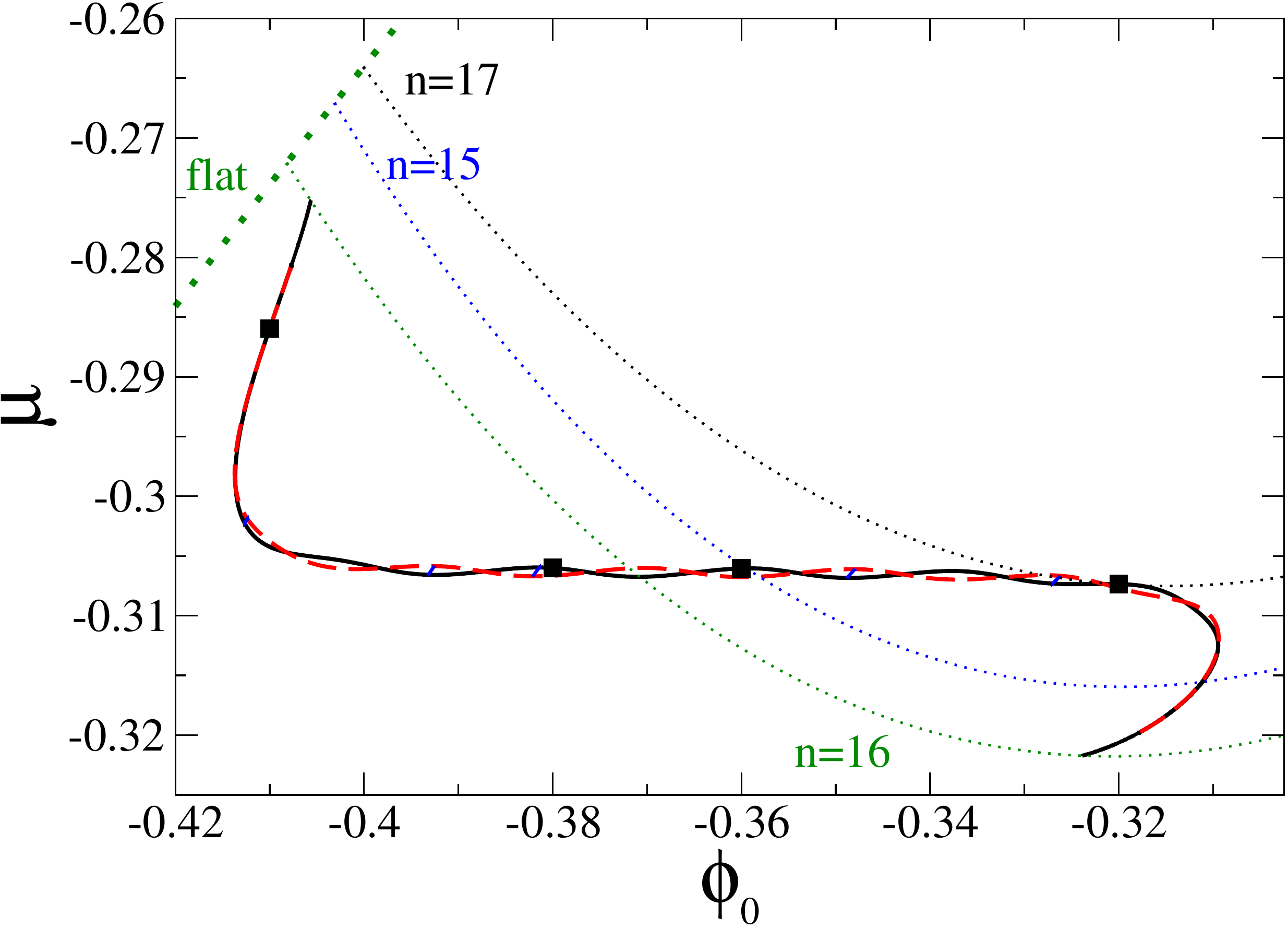}\textbf{(f)}
\textbf{(g)}\includegraphics[width=0.425\hsize]{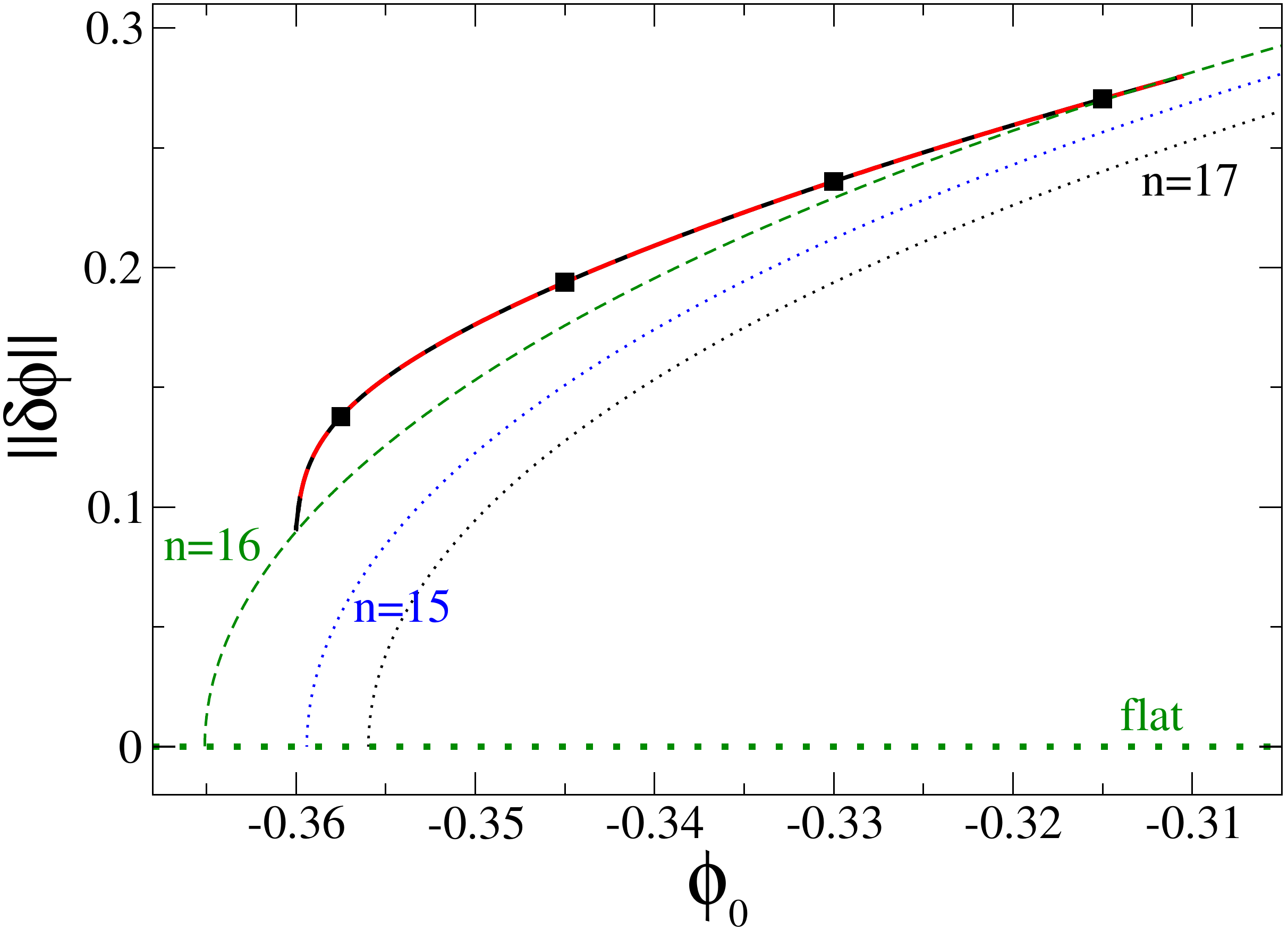}
\includegraphics[width=0.425\hsize]{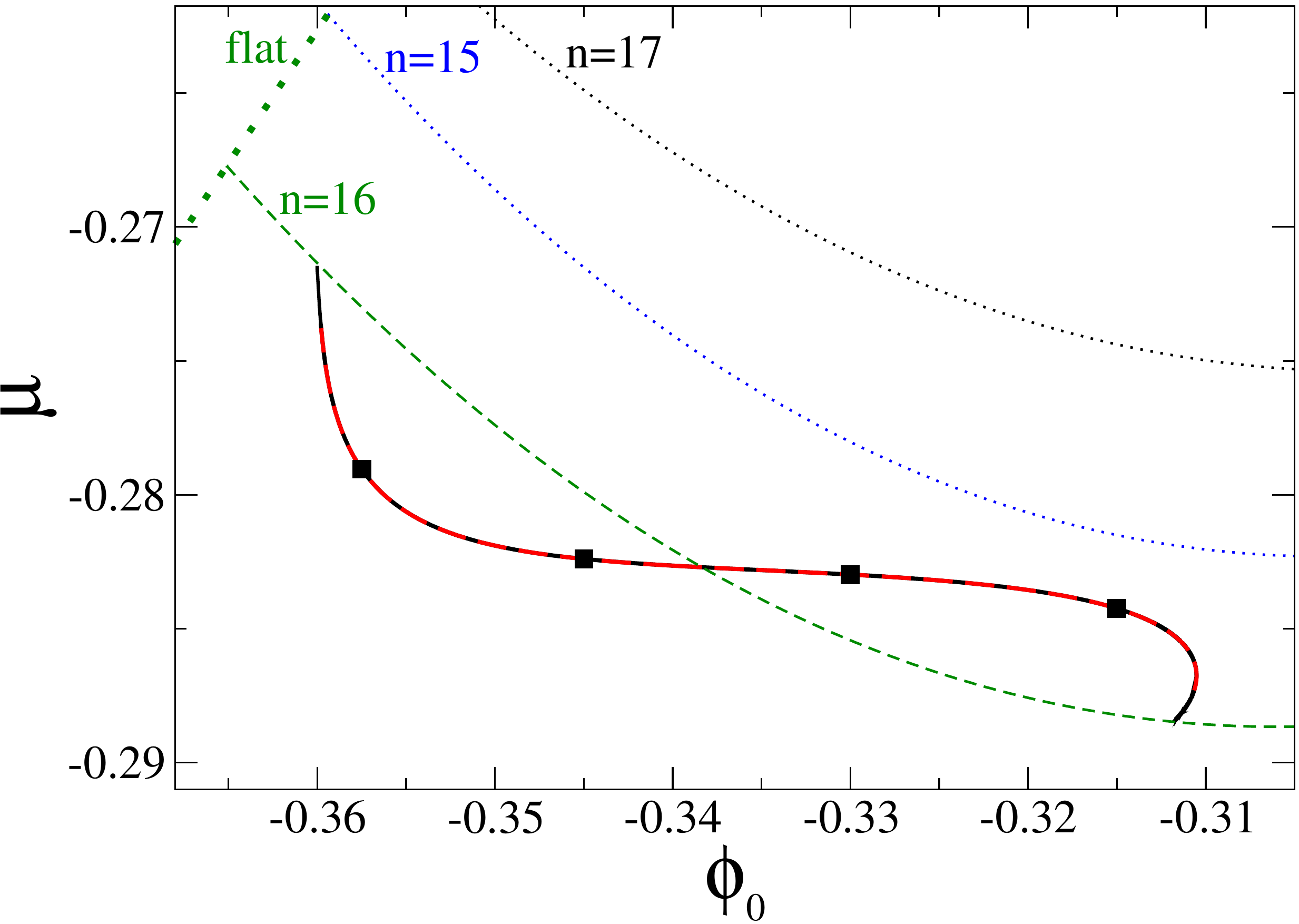}\textbf{(h)}
\caption{(color online) The norm (left) and chemical
  potential (right) of the homogeneous, periodic and localized steady state 
  solutions as a function of the mean order parameter $\phi_0$, for a fixed 
  domain size of $L=100$ and several values of $r>-0.9$. In (a,b) $r=-0.7$, 
  (c,d) $r=-0.6$, (e,f) $r=-0.5$, and (g,h)
  $r=-0.4$.  The line styles are as in Fig.~\ref{fig:loc-fam-rm09}. 
  Typical order parameter profiles 
  along the branches of symmetric localized states with an odd
  number of maxima (black lines) are shown in
  Fig.~\ref{fig:loc-prof-rm07-0375}, and correspond to locations 
  indicated in the panels by filled black squares.
}
\mylab{fig:loc-fam-sevrm-one}
\end{figure}

\begin{figure}
	\includegraphics[width=0.9\hsize]{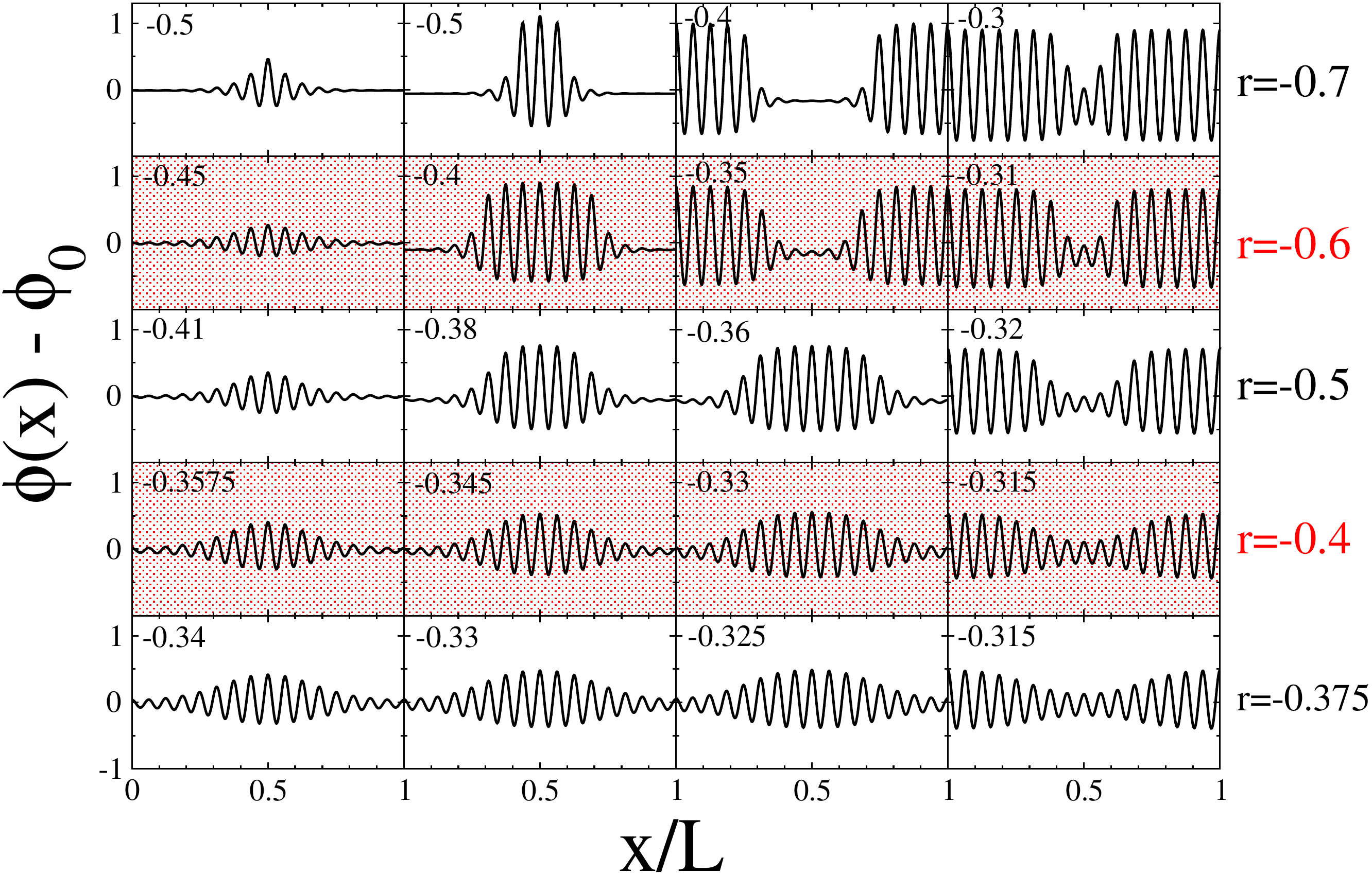}
	\caption{(Color online) A selection of steady state profiles $\phi(x)-\phi_0$
          along the LS$_\mathrm{odd}$ branches shown in 
          Fig.~\ref{fig:loc-fam-sevrm-one} for $r=-0.7$ (1st row), $r=-0.6$
          (2nd row), $r=-0.5$ (3rd row), $r=-0.4$ (4th row), and
          $r=-0.375$ (last row), for various values of ${\phi}_0$, as indicated in
          the top left corner of each panel. The solutions in each row
          are shown in order, starting from near
          the bifurcation point that creates them and continuing to larger 
          norm $||\delta \phi||$. The locations of the profiles are indicated
          in Fig.~\ref{fig:loc-fam-sevrm-one} by the filled
          black squares. The bifurcation diagram for $r=-0.375$ (not shown
          in Fig.~\ref{fig:loc-fam-sevrm-one}) is
          qualitatively the same as that for $r=-0.4$.
        }
	\label{fig:loc-prof-rm07-0375}
\end{figure}

We now show how the bifurcation diagrams evolve as the temperature-like
parameter $r$ changes. We begin by showing the bifurcation diagrams
for decreasing values of $|r|$. In the Appendix we use amplitude
equations to determine the direction of branching of the localized
states. Here we discuss the continuation results.

The bifurcation diagram for $r=-0.7$
(Fig.~\ref{fig:loc-fam-sevrm-one}(a,b)) resembles that for $r=-0.9$
(Fig.~\ref{fig:loc-fam-rm09}) although the snaking structure has moved
towards smaller $|\phi_0|$ and is now thinner. In addition, it is now
the second saddle-node on the LS$_\mathrm{odd}$ branch that lies
farthest to the left, and not the first. For $r=-0.6$
(Fig.~\ref{fig:loc-fam-sevrm-one}(c,d)), the branches of localized
states still form a tilted snakes-and-ladders structure, but the
saddle-nodes on the LS$_\mathrm{odd}$ and LS$_\mathrm{even}$ branches
are now absent, i.e., both solution branches now grow
monotonically. The resulting diagram has been called ``smooth
snaking'' \cite{DL10}. However, despite the absence of the
saddle-nodes on the LS$_\mathrm{odd}$ and LS$_\mathrm{even}$ branches
the interconnecting ladder states consisting of asymmetric states
still remain. This continues to be the case when $r=-0.5$
(Fig.~\ref{fig:loc-fam-sevrm-one}(e,f)) although the structure has
moved to yet larger $\phi_0$ and the snake has become even
thinner. Finally, for $r=-0.4$ (Fig.~\ref{fig:loc-fam-sevrm-one}(g,h)),
the snake is nearly dead, and only tiny wiggles remain. The
bifurcation of the localized states from the $n=16$ periodic state is
now supercritical (see Appendix) but the LS branches continue to 
terminate on the
same branch at larger amplitude, and do so via a single saddle-node at
the right (Fig.~\ref{fig:loc-fam-sevrm-one}(h)). Sample
profiles along the resulting LS$_\mathrm{odd}$ branches are shown
for several values of $r$ in Fig.~\ref{fig:loc-prof-rm07-0375}.  We
note that a change of $r$ has a profound effect on the transition
region between the homogeneous background state and the periodic
state: with decreasing $|r|$ the LS become wider and the localized
periodic structure looks more and more like a wave packet with a smooth
sinusoidal modulation of the peak amplitude.

As the `temperature' $r$ decreases even further, the bifurcation
diagrams remain similar to those displayed in Fig.~\ref{fig:loc-fam-rm09} 
until just before
$r=-1.5$, where substantial changes take place and the complexity of the 
bifurcation diagram grows dramatically. This is a consequence of the 
appearance of other types of localized states that we do not discuss here. 
Likewise, we omit here all bound states of the LS described above. These are
normally found on an infinite stack of isolas that are also present in 
the snaking region \cite{bukn09,KLSW:11}.

\subsection{Tracking the snake}
\label{sec:snake}

In Fig.~\ref{fig:prof-loc-folds} we show, for $-1.5<r<-0.4$, the 
result of tracking \textit{all} the saddle-node bifurcations visible in 
the previous bifurcation diagrams in the $(\phi_0,r)$ plane, while 
Fig.~\ref{fig:prof-loc-folds-zoom} shows an enlargement of the region
$-1<r<-0.35$ together with the result of tracking the tertiary
pitchfork bifurcations to the asymmetric states.

Figure \ref{fig:prof-loc-folds} shows that the saddle-nodes annihilate
pairwise in cusps as $r$ increases. The annihilations occur first for 
smaller $\phi_0$ and later for larger $\phi_0$, and occur alternately on
LS$_\mathrm{odd}$ and LS$_\mathrm{even}$. Above the locus of the cusps the 
snaking is smooth, although as shown in Fig.~\ref{fig:prof-loc-folds-zoom} 
the tertiary ladder states remain. The thick green curve in
Fig.~\ref{fig:prof-loc-folds-zoom} represents the locus of the secondary 
bifurcation from the $n=16$ periodic state to LS and shows that on either 
side the bifurcation to LS is subcritical for sufficiently negative $r$ but 
becomes supercritical at larger $r$, cf.~Fig.~\ref{fig:loc-fam-sevrm-one}(h) 
and Appendix.

\begin{figure}
\includegraphics[width=0.9\hsize]{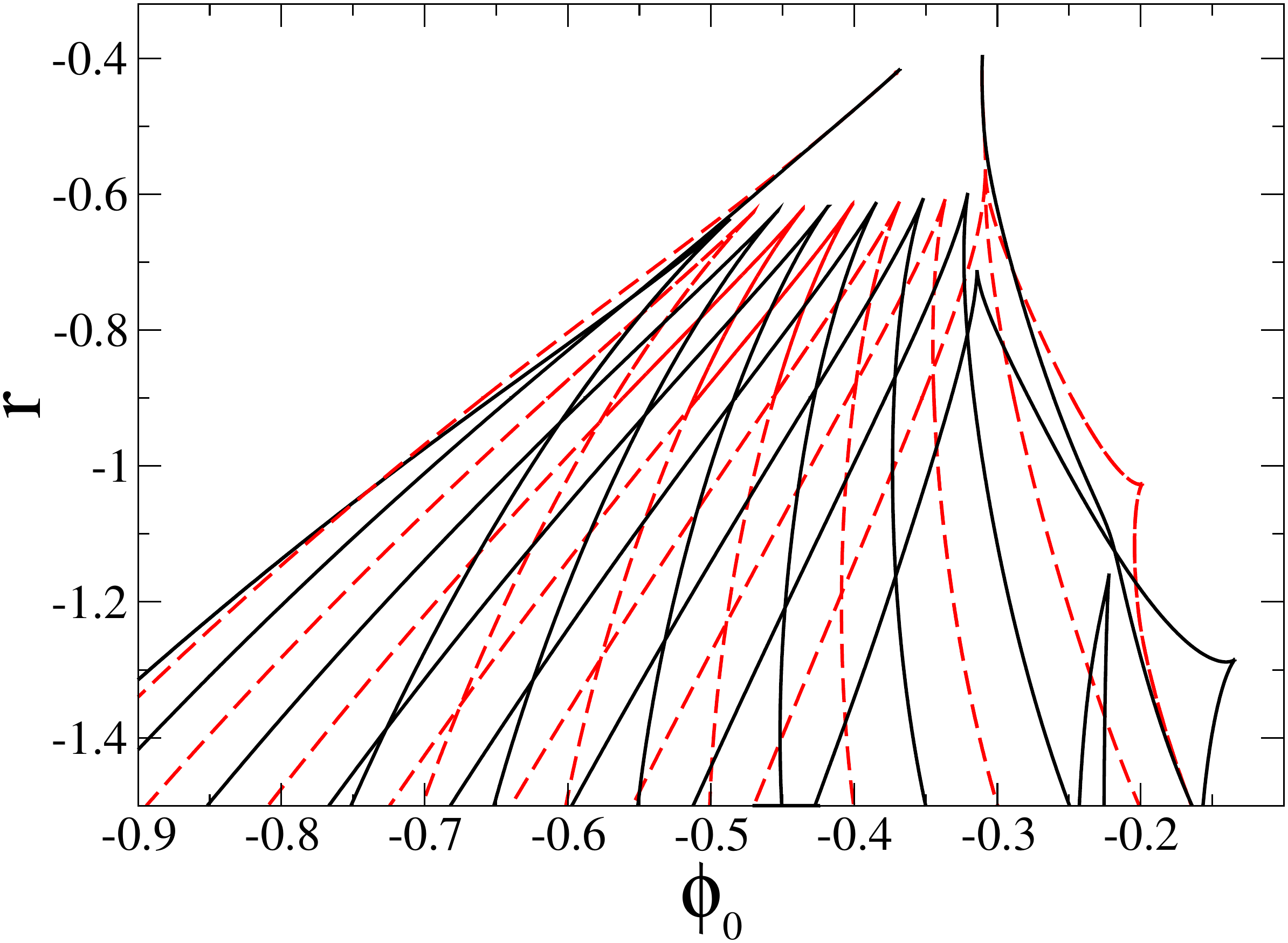}  
\caption{(color online) Loci of saddle-node bifurcations on the branches of symmetric 
localized states in the $(\phi_0,r)$ plane for $r>-1.5$. Saddle-nodes 
annihilate pairwise as $r$ increases (solid black lines: LS$_\mathrm{odd}$; 
dashed red lines: LS$_\mathrm{even}$).
}
\mylab{fig:prof-loc-folds}
\end{figure}

\begin{figure}
\includegraphics[width=0.9\hsize]{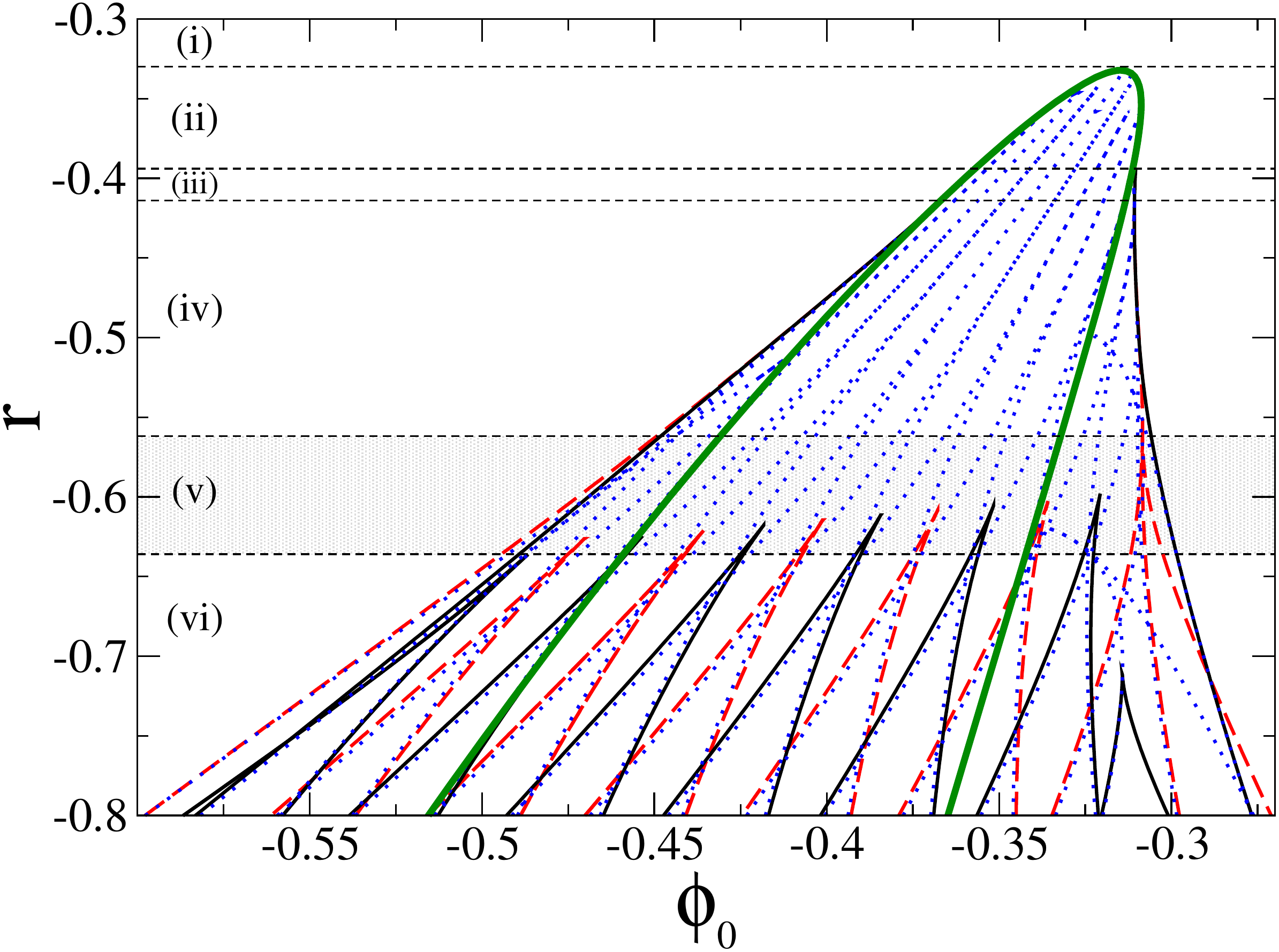}  
\caption{(color online) Loci of the saddle-node bifurcations on the branches of symmetric 
localized states in the $(\phi_0,r)$ plane (solid black lines: LS$_\mathrm{odd}$; 
dashed red lines: LS$_\mathrm{even}$) together with the bifurcations to
asymmetric localized states (dotted blue lines) for $r>-1.5$. The thick green curve
represents the locus of secondary bifurcation from the periodic $n=16$
state to localized states and shows that
on either side the bifurcation to LS is subcritical for sufficiently
negative $r$ but becomes supercritical at larger $r$.
}
\mylab{fig:prof-loc-folds-zoom}
\end{figure}

One may distinguish six intervals in $r$ with different types of 
behavior. These depend on the system size, but the results in 
Fig.~\ref{fig:prof-loc-folds-zoom} for $L=100$ are representative:
\begin{itemize}
\item[(i)] For $r>-0.33$: no LS exist, the only nontrivial states are
  periodic solutions.
\item[(ii)] For $-0.33>r>-0.39$: branches of even and odd symmetric LS
  are present, and appear and disappear via supercritical secondary bifurcations
  from the branch of periodic solutions. With decreasing $r$, more and more
  branches of asymmetric LS emerge from these two secondary bifurcation points.
\item[(iii)] For $-0.39>r>-0.41$: both branches of symmetric LS emerge
  subcritically at large $\phi_0$ and supercritically at small $\phi_0$.
\item[(iv)] For $-0.41>r>-0.56$: both branches of symmetric LS emerge
  subcritically at either end. Further branches of asymmetric LS
  emerge with decreasing $r$ either from the two secondary bifurcation
    points or from the saddle-node bifurcations on the branches of
    symmetric LS, but the symmetric LS still do not exhibit snaking,
    i.e., no additional folds are present on the branches of
  symmetric LS.
\item[(v)] For $-0.56>r>-0.64$ (highlighted by the grey shading in
  Fig.~\ref{fig:prof-loc-folds-zoom}): pairs of saddle-nodes
  appear successively in cusps as $r$ decreases, starting at
  larger $\phi_0$. Thereafter saddle-nodes appear alternately on branches of 
  even and odd symmetric LS. The appearance of the cusps is therefore 
  associated with the transition from smooth snaking to slanted snaking.
\item[(vi)] For $-0.64>r$: The slanted snake is fully developed. Only one further
  pair of saddle-node bifurcations appears in the parameter region
  shown in Fig.~\ref{fig:prof-loc-folds-zoom}. With decreasing $r$ the
  snaking becomes stronger; each line in
  Figs.~\ref{fig:prof-loc-folds} and \ref{fig:prof-loc-folds-zoom}
  that represents a saddle-node bifurcation crosses more and more
  other such lines, i.e., more and more different states are possible
  at the same values of $\phi_0$. Furthermore, the subcritical regions 
  (outside the green curve in Fig.~\ref{fig:prof-loc-folds-zoom}) become larger.
\end{itemize}

\subsection{Relation to binodal lines}

From the condensed matter point of view, where the cSH/PFC equation
represents a model for the liquid (homogeneous) and solid (periodic) phases, 
one is particularly interested in results in the thermodynamic limit 
$L\to \infty$. As mentioned above in the context of the phase diagram in 
Fig.\ \ref{fig:cSH-phasediagram-1d}, the binodal lines correspond to 
values of $(\phi_0,r)$ at which the homogeneous state and the minimum 
energy periodic state coexist in the thermodynamic limit. These are 
defined as pairs of points at which the homogeneous state and the periodic
state have the same `temperature' (i.e., same $r$ value), the same chemical 
potential $\mu$ and the same pressure $p=-\omega$, and are displayed as the 
blue dash-dot lines in Fig.~\ref{fig:prof-loc-folds-binodal}. For a given 
value of $r$, these two lines give the values of $\phi_0$ of the coexisting 
homogeneous (lower $\phi_0$) and periodic (higher $\phi_0$) states. Note 
that when plotted with the resolution of Fig.~\ref{fig:prof-loc-folds-binodal},
the binodals are indistinguishable from the coexistence lines between the 
finite size $L=100$, $n=16$ periodic solution and the homogeneous state. 
Figure~\ref{fig:prof-loc-folds-binodal} also displays the line (green solid 
line) at which the $L=100$ localized states bifurcate from the $n=16$ branch 
of periodic solutions. For $r<-0.4$ the LS bifurcations
are subcritical implying that the localized states are present outside the
green solid line. Figure~\ref{fig:prof-loc-folds-binodal} shows the loci of 
the outermost saddle-node bifurcations on the branches of symmetric LS that 
result (dashed lines to the left and to the right of the green solid line for 
$r<-0.4$). The most striking aspect of Fig.~\ref{fig:prof-loc-folds-binodal}
is that for $r\lesssim-1$ these lines actually cross and exit the region 
between the two binodal lines, indicating that in the PFC model one can find 
stable LS {\em outside} of the binodal. Although these are not the lowest 
free energy states (we have checked this for $r\gtrsim-1.5$), this remarkable 
fact points towards the possibility of metastable nanocrystals existing 
outside of the binodal. We must mention, however, that these structures have 
been found in a finite size system with $L=100$; we have not investigated 
their properties for larger system sizes $L$. 

\begin{figure}
\includegraphics[width=0.9\hsize]{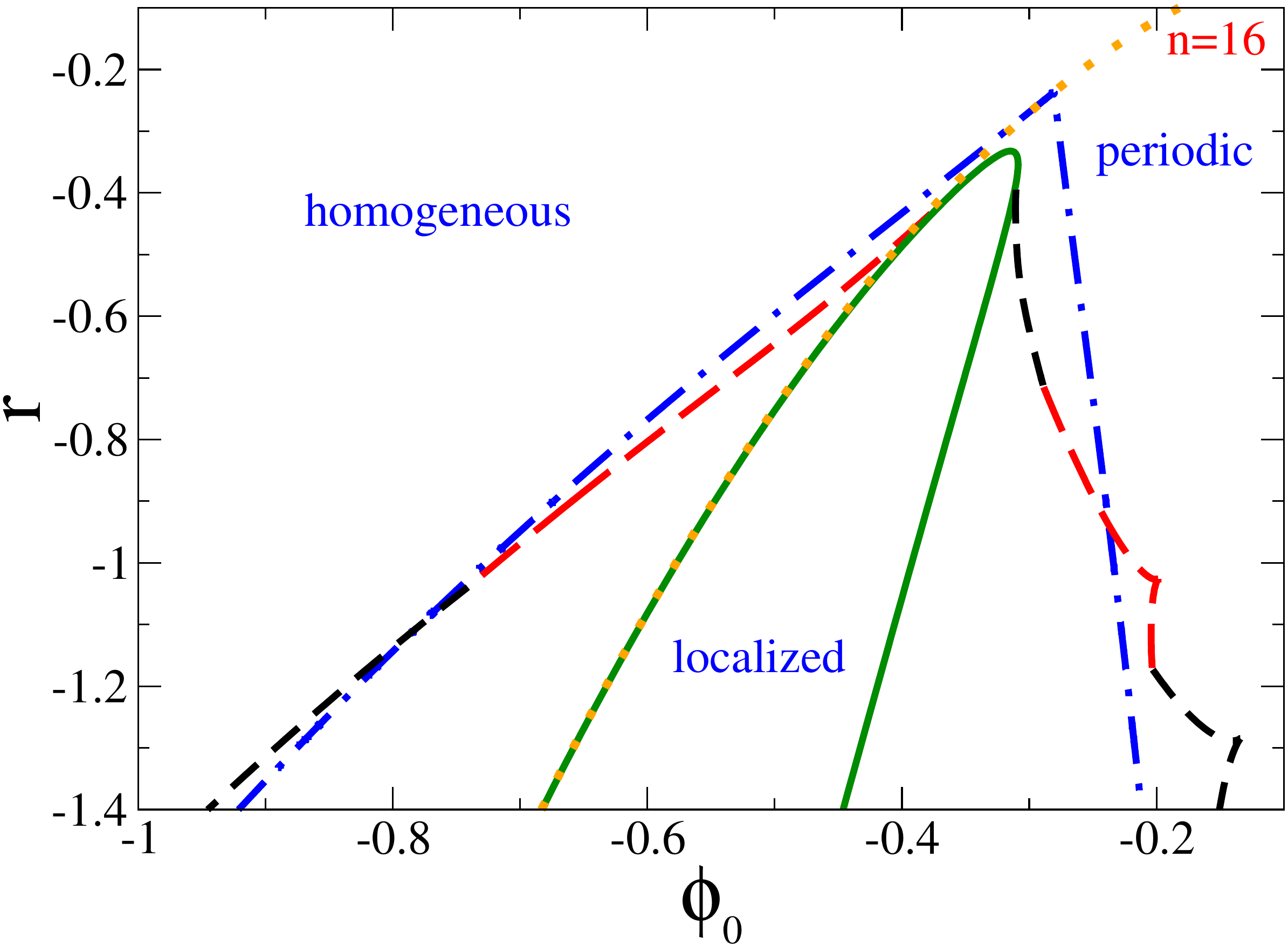}
\caption{(color online) Loci in the $(\phi_0,r)$ plane computed for $L=100$
of (i) the primary bifurcation from the homogeneous state to the $n=16$ 
periodic branch (dotted orange line); (ii) the bifurcation of the
localized states from the $n=16$ periodic branch (solid green line);
(iii) the outermost saddle-node bifurcations on the branches of
symmetric localized states (long-dashed red and short-dashed black
lines); and (iv) the binodals between the periodic and homogeneous 
states (dot-dashed blue lines). The latter coincide with the binodals
for the $n=16$ periodic state and the homogeneous state to within the
resolution of the figure. 
Figures~\ref{fig:prof-loc-folds} and \ref{fig:prof-loc-folds-zoom}
show how the loci of the outermost saddle-node bifurcations of
localized states fit into the overall picture.}
  \mylab{fig:prof-loc-folds-binodal}
\end{figure}

\subsection{Stability}

Although we have not computed the stability properties of the different 
localized states with respect to infinitesimal perturbations, we can use 
the principles of bifurcation theory to deduce the likely stability 
properties. In the reference bifurcation diagram in 
Fig.~\ref{fig:loc-fam-rm09}(a) the homogeneous state (liquid) is stable for 
large negative $\phi_0$, and for a system with $L=100$ loses stability 
to the $n=16$ mode as $\phi_0$ increases. Since the bifurcation is 
supercritical, the $n=16$ periodic state is initially stable. The LS 
that bifurcate from it subcritically will both be unstable. The single 
bump state is likely to be once unstable and it therefore acquires 
stability at the first saddle-node on the left, remaining stable until 
the first crosslink where a second (phase) eigenvalue becomes unstable. 
The first (amplitude) eigenvalue becomes unstable at the saddle-node on 
the right so that the portion of LS$_\mathrm{odd}$ with negative slope is 
twice unstable. These instabilities are then undone so that the 
LS$_\mathrm{odd}$ above the second crosslink are again stable. The 
transition to smooth snaking that occurs with decreasing $r$ eliminates 
instabilities associated with the first eigenvalue but not the second.

The LS$_\mathrm{even}$ states are initially twice unstable but both eigenvalues 
stabilize near the first saddle-node on the left, so that LS$_\mathrm{even}$
is stable on the part of the branch with positive slope, below the second
crosslink etc. Thus the connecting LS$_\mathrm{asym}$ are unstable. Once again, 
the transition to smooth snaking that occurs with decreasing $r$ eliminates 
instabilities associated with the amplitude eigenvalue but not those associated 
with the phase eigenvalue. One can check that in this case the connecting 
LS$_\mathrm{asym}$ remain unstable with the LS$_\mathrm{odd}$ stable between the
second and third crosslinks (Fig.~\ref{fig:loc-fam-sevrm-one}(c)), and unstable 
between the third and fourth crosslinks. Likewise the LS$_\mathrm{even}$ are 
stable above the left saddle-node and below the second crosslink; they are 
unstable between the second and third crosslinks and acquire stability at the 
third crosslink etc (Fig.~\ref{fig:loc-fam-sevrm-one}(c)). Note that as a 
result of these stability assignments there is at least one LS state that is 
stable at all $\phi_0$ values between the leftmost and rightmost saddle-nodes.

\section{Localized states in two and three dimensions}
\mylab{sec:2d3d}

\subsection{Numerical algorithm}

To perform direct numerical simulations (DNS) of the conserved Swift-Hohenberg 
equation in higher dimensions we use a recently proposed algorithm 
\cite{GoNo12} that has been proved to be unconditionally energy-stable. As a 
consequence, the algorithm produces free-energy-decreasing discrete solutions,
irrespective of the time step and the mesh size, thereby respecting
the thermodynamics of the model even for coarse
discretizations. For the spatial discretization we employ Isogeometric
Analysis \cite{HCB05}, which is a generalization of the Finite Element
Method. The key idea behind Isogeometric Analysis is the use
of Non-Uniform Rational B-Splines (NURBS) instead of the standard
piecewise polynomials used in the Finite Element Method. With
NURBS Isogeometric Analysis gains several advantages over the
Finite Element Method. In the context of the conserved Swift-Hohenberg
equation, the most relevant one is that the Isogeometric Analysis permits
the generation of arbitrarily smooth basis functions that lead to a
straightforward discretization of the higher-order partial derivatives
of the conserved Swift-Hohenberg equation \cite{GCBH08}. For the time
discretization we use an algorithm especially designed for the
conserved Swift-Hohenberg equation. It may be thought of as a
second-order perturbation of the trapezoidal rule which achieves
unconditional stability, in contrast with the trapezoidal scheme. All
details about the numerical algorithms may be found in~\cite{GoNo12}.

\subsection{Two dimensions}

\begin{figure}
\begin{minipage}[b]{0.20\hsize}
\includegraphics[width=\hsize]{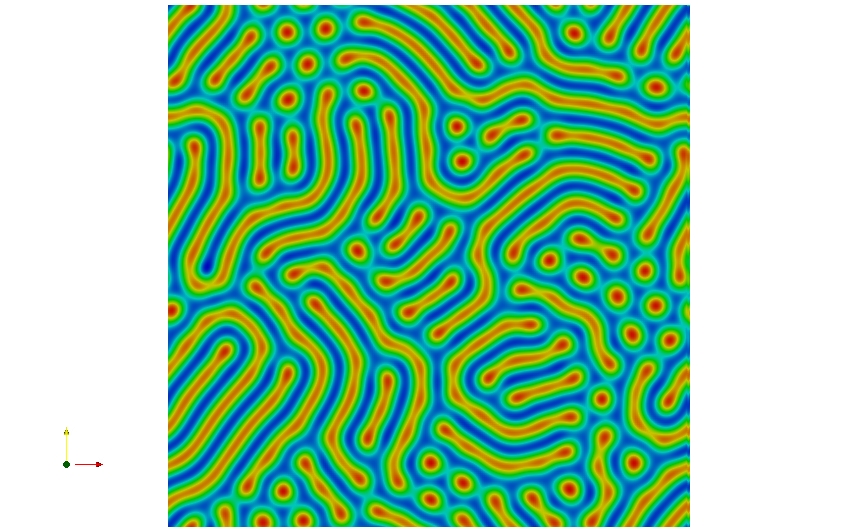}
\includegraphics[width=\hsize]{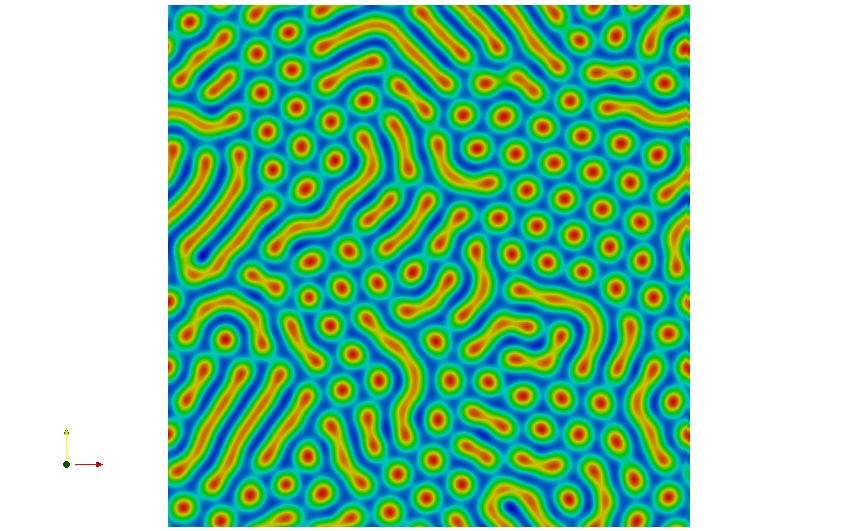}
\includegraphics[width=\hsize]{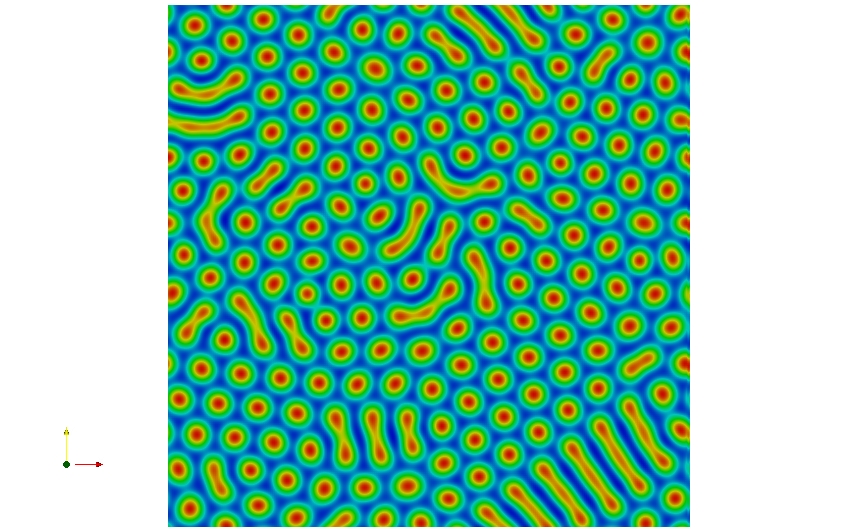}\\ \phantom{\small{xx}}
\end{minipage}
\includegraphics[width=0.55\hsize]{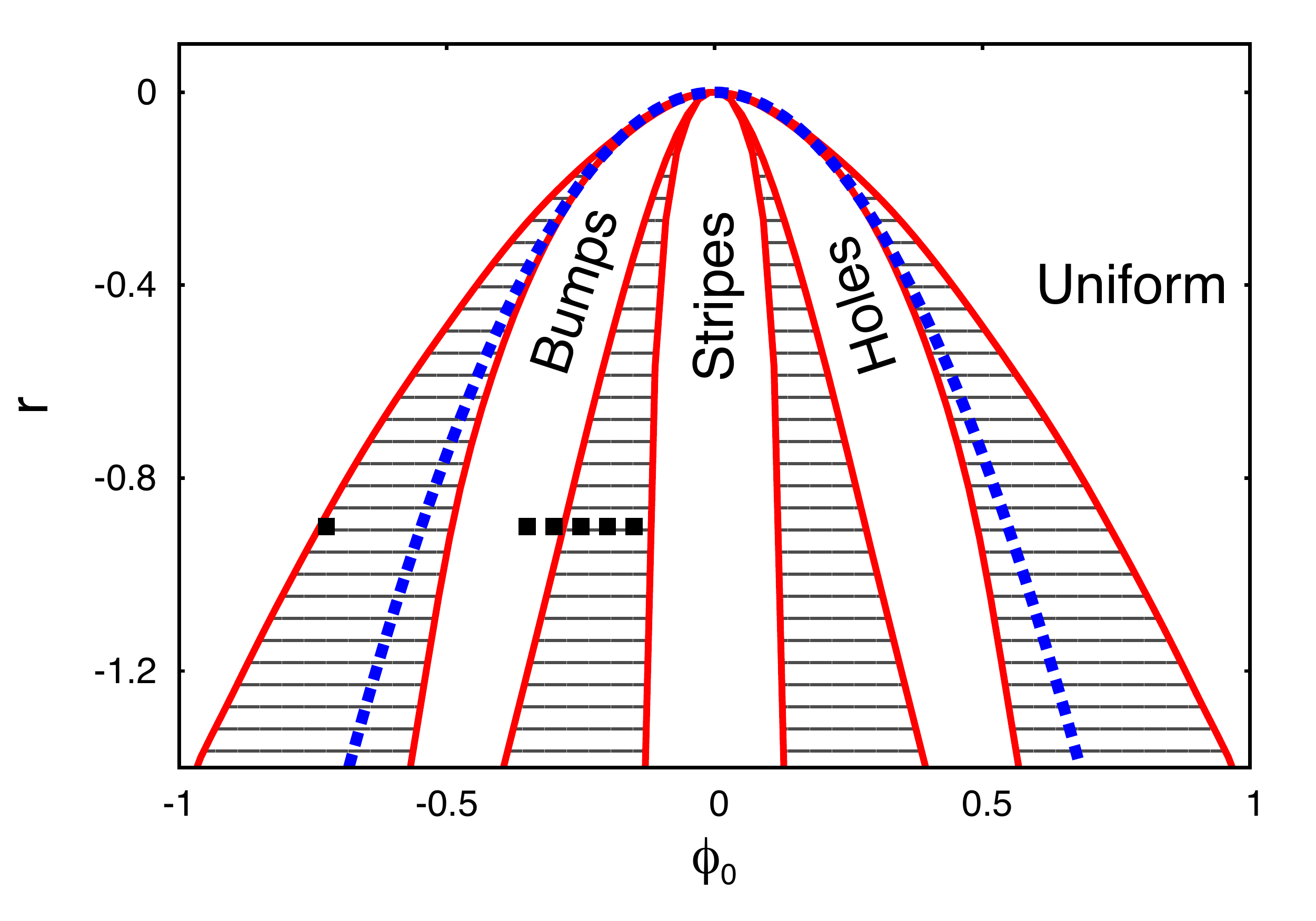}
\begin{minipage}[b]{0.20\hsize}
\includegraphics[width=\hsize]{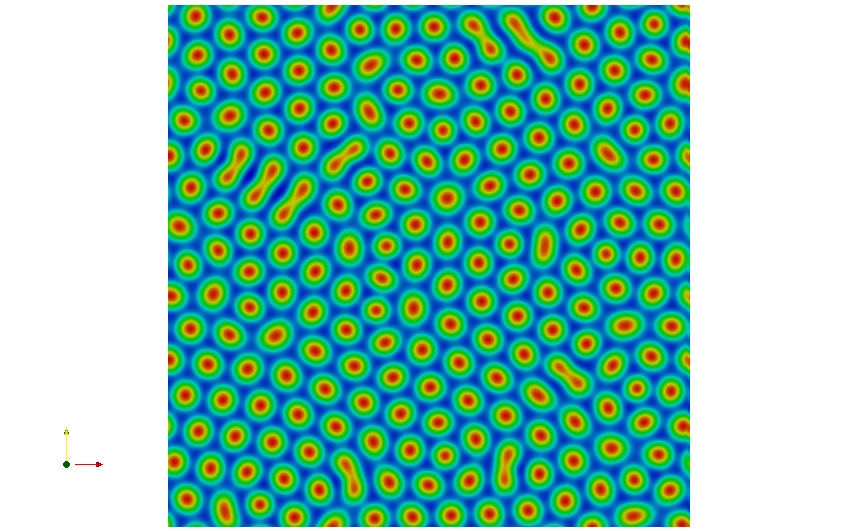}
\includegraphics[width=\hsize]{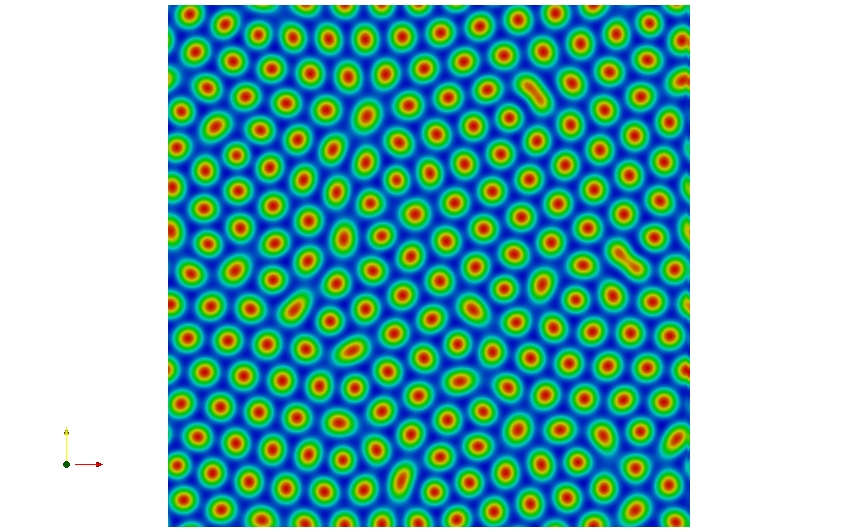}
\includegraphics[width=\hsize]{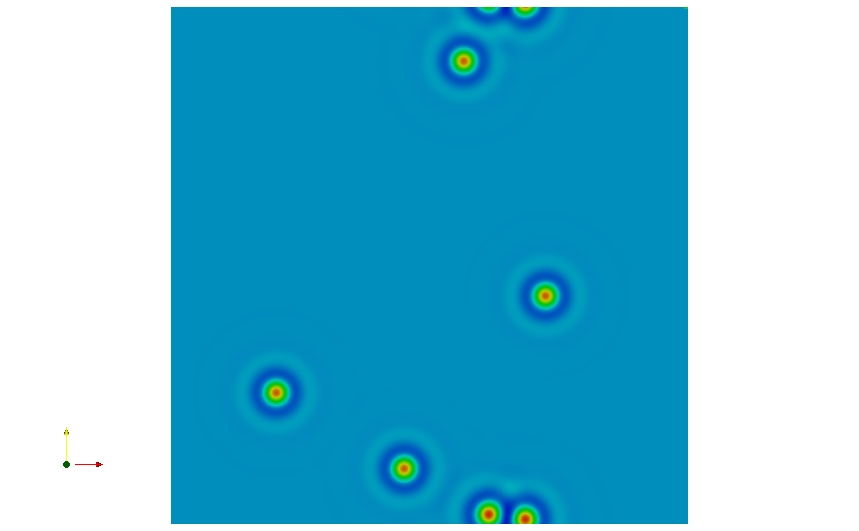}\\ \phantom{\small{xx}}
\end{minipage}
\caption{(color online) Phase diagram for the conserved
  Swift-Hohenberg equation, Eq.~(\ref{eq:csh}), in two dimensions when
  $q = 1$. The red solid lines show the various coexistence curves, 
  the blue dotted line shows the limit of linear stability and the grey
  striped areas show phase coexistence regions. The small panels show
  steady solutions at $r=-0.9$ and selected mean order parameter values 
  $\phi_0$ indicated by the black squares. From top
  to bottom left $\phi_0=-0.15$, $-0.2$ and $-0.25$, while from top to
  bottom right $\phi_0=-0.3$, $-0.35$ and $-0.725$.  Additional
  solutions that detail the transition occurring between the last two
  panels are shown in Fig.~\ref{fig:prof-twodim-one}. The domain size
  is $100\times100$.}
\mylab{fig:cSH-phasediagram-2d}
\end{figure}

As in one spatial dimension, the phase diagram for two-dimensional structures
helps us to identify suitable parameter values where LS are likely to occur. 
The phase diagram in Fig.~\ref{fig:cSH-phasediagram-2d}, determined numerically 
\cite{ARTK12}, shows three distinct phases,
labeled `bumps', `stripes' and `holes'. In view of the fact that 
$\phi(x)=\phi_0+\delta\phi(x)$, bumps and holes correspond to perturbations
$\delta\phi(x)$ with opposite signs; both have hexagonal coordination. In 
addition, the phase diagram reveals four regions of thermodynamic
coexistence (hatched in Fig.\ref{fig:cSH-phasediagram-2d}), 
between bumps and the uniform state, between bumps and stripes, between 
stripes and holes, and between holes and the uniform state, respectively.

Examples of results obtained by DNS of Eq.~(\ref{eq:csh}) starting
from random initial conditions are displayed in the side panels of
Fig.~\ref{fig:cSH-phasediagram-2d}. These six panels show results for
fixed $r=-0.9$. For small values of $|\phi_0|$ the system forms a
labyrinthine lamellar-like stripe state; the stripes pinch off locally
into bumps as $|\phi_0|$ is increased, leading to the formation of
inclusions of bumps in a background stripe state. The pinching tends
to occur first at the ends of a stripe and then proceeds gradually
inwards. In other cases, free ends are created by the splitting of a
stripe into two in a region of high curvature. The formation of bumps
tends to take place at grain boundaries, and once bump formation
starts, it tends to spread outward from the initial site.  For
$\phi_0=-0.2$ the areas covered by stripes and bumps are comparable
and for larger values of $|\phi_0|$ the bump state dominates. By
$\phi_0=-0.3$ the stripes are almost entirely gone and the state takes
the form of a crystalline solid with hexagonal coordination but having
numerous defects. As $|\phi_0|$ increases further, vacancies appear in
the solid matrix and for large enough $|\phi_0|$ the solid ``melts'' into
individual bumps or smaller clusters, as further detailed in
Fig.~\ref{fig:prof-twodim-one}.

Figure \ref{fig:prof-twodim-one} shows further results from a scan through
decreasing values of $\phi_0$ at $r=-0.9$. We focus on the relatively
small range $\phi_0=-0.45$ to $\phi_0=-0.675$, where localized states
occur. These reveal a gradual transition from a densely packed solid-like
structure to states with a progressively increasing domain area that is 
free of bumps, i.e., containing the homogeneous state.  The bumps 
percolate through the domain until approximately $\phi_0=-0.575$. For 
smaller values of $\phi_0$ the order parameter profiles resembles a 
suspension of solid fragments in a liquid phase; the solid is no longer
connected. As $\phi_0$ decreases further the characteristic size of
the solid fragments decreases, as the solid fraction falls.

\begin{figure}
\includegraphics[width=1.0\hsize]{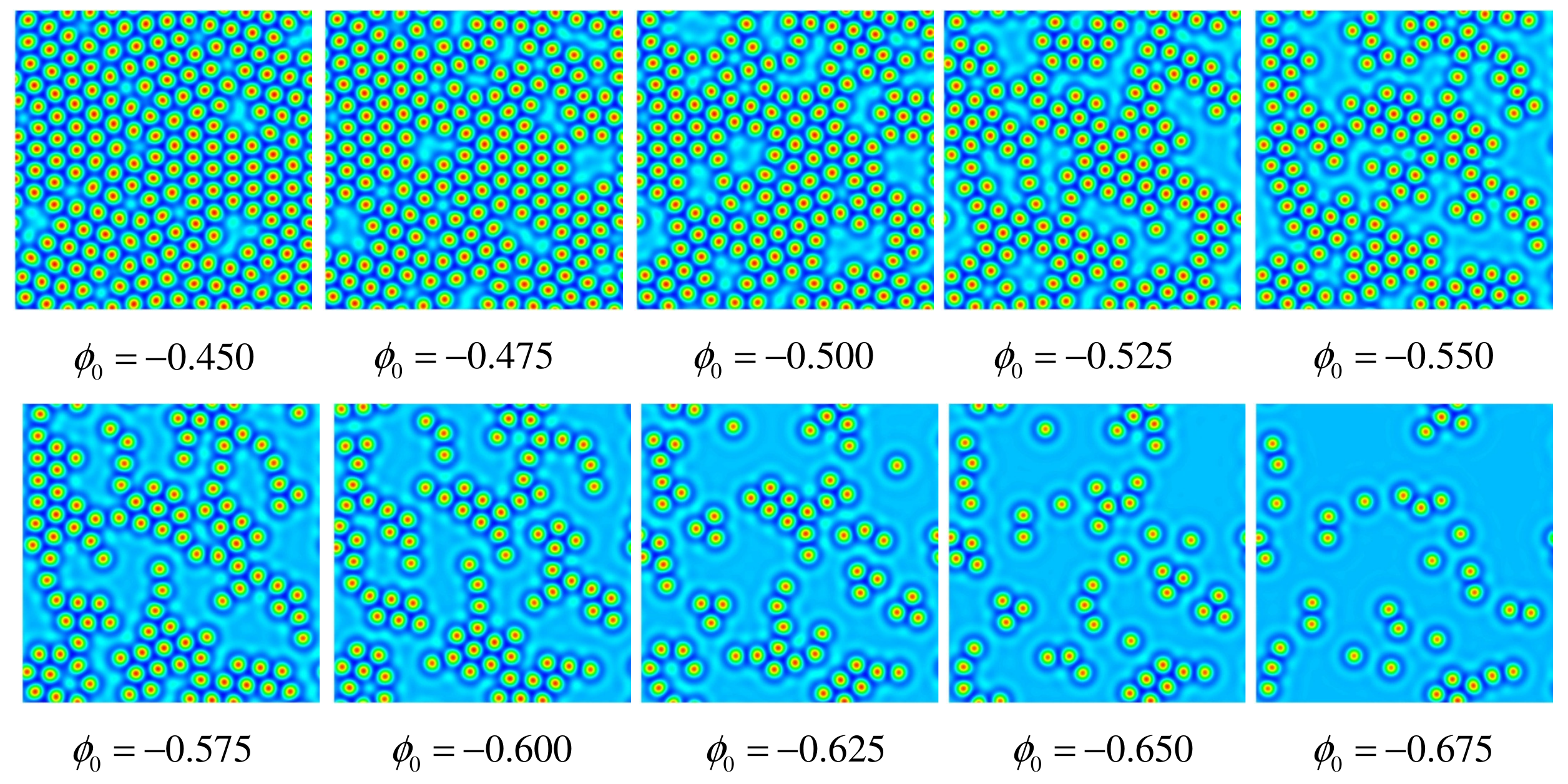}
\caption{(color online) Steady-state solutions of the conserved
  Swift-Hohenberg equation, Eq.~(\ref{eq:csh}), in two dimensions for
  $r=-0.9$ and different values of $\phi_0$ in the
  range $-0.675<\phi_0<-0.45$, where localized states occur. 
  The corresponding value of $\phi_0$ is indicated below
  each panel. The domain size is $100\times100$.}
\mylab{fig:prof-twodim-one}
\end{figure}

\subsection{Three dimensions}

In three dimensions, Eq.~(\ref{eq:csh}) exhibits a large number of steady state
spatially periodic
structures. These include those with the symmetries of the simple cubic lattice,
the face-centered cubic lattice and the body-centered cubic lattice 
\cite{Callahan}. Although we do not calculate the phase diagram for the
three-dimensional (3d) 
system, numerical simulations in 3d reveal that a lamellar (parallel `sheets')
state is energetically preferred for small $\phi_0$. Slices through these 
structures resemble the stripes observed in 2d. As $|\phi_0|$ increases, the 
lamellae pinch off, much as in two dimensions, and progressively generate a 
3d disordered array of bumps (Fig.~\ref{fig:prof-threedim-one}). This 
solid-like state is far from being a perfect crystal, however, and with 
increased $|\phi_0|$ develops vacancies which eventually lead to its 
dissolution, just as in two dimensions.

\begin{figure}
\includegraphics[width=0.32\hsize]{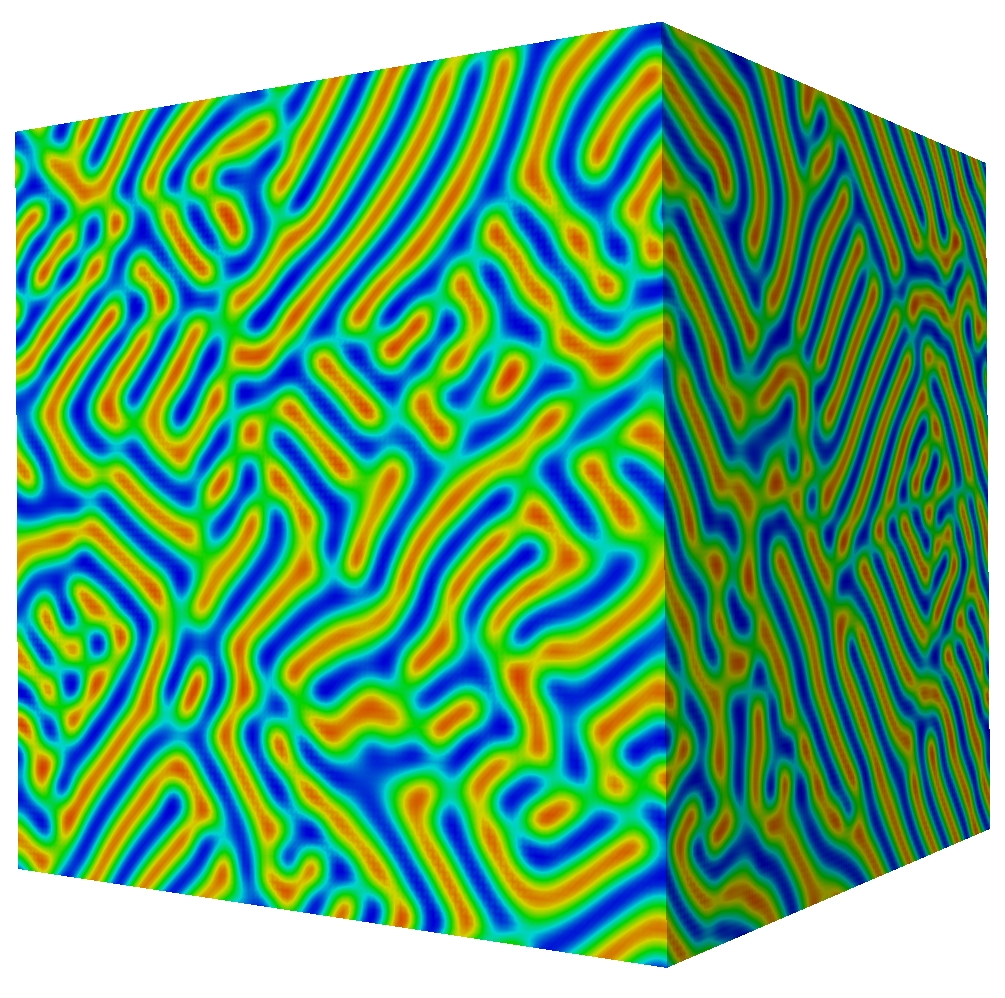}
\includegraphics[width=0.32\hsize]{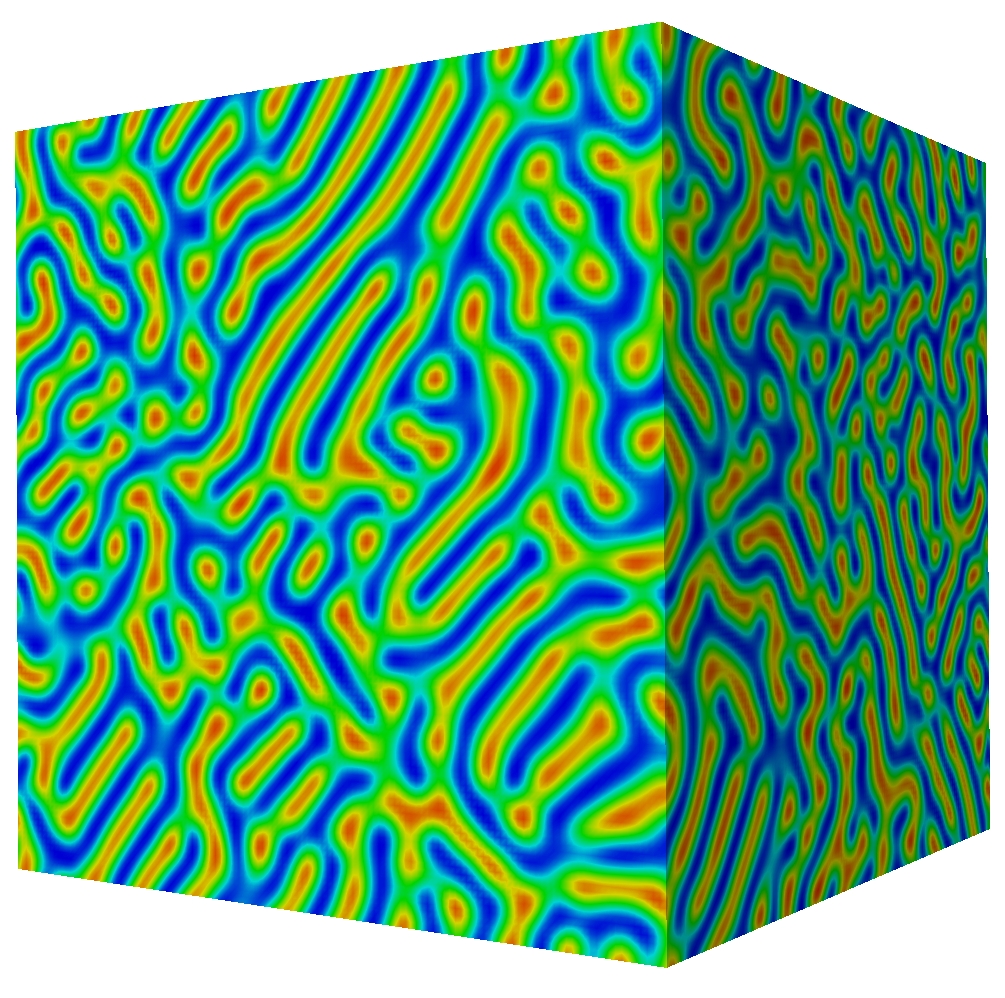}
\includegraphics[width=0.32\hsize]{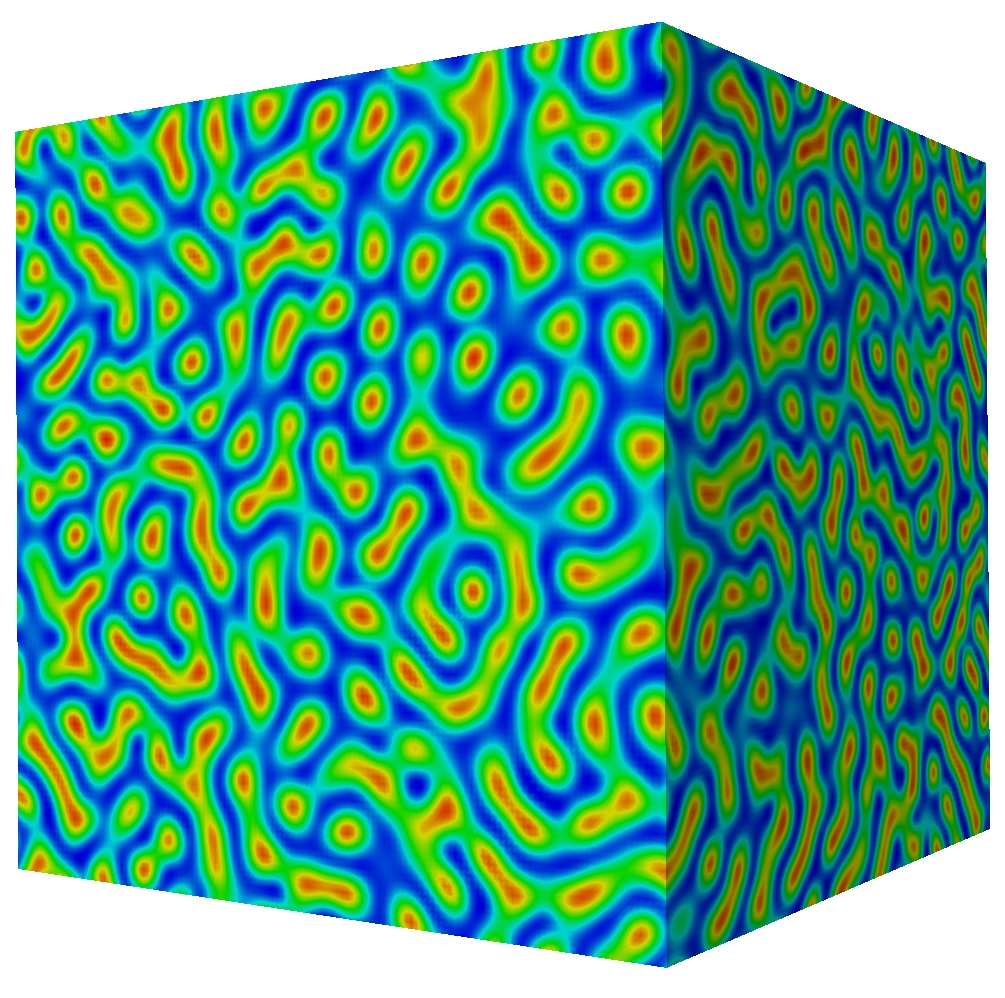}\\
\includegraphics[width=0.32\hsize]{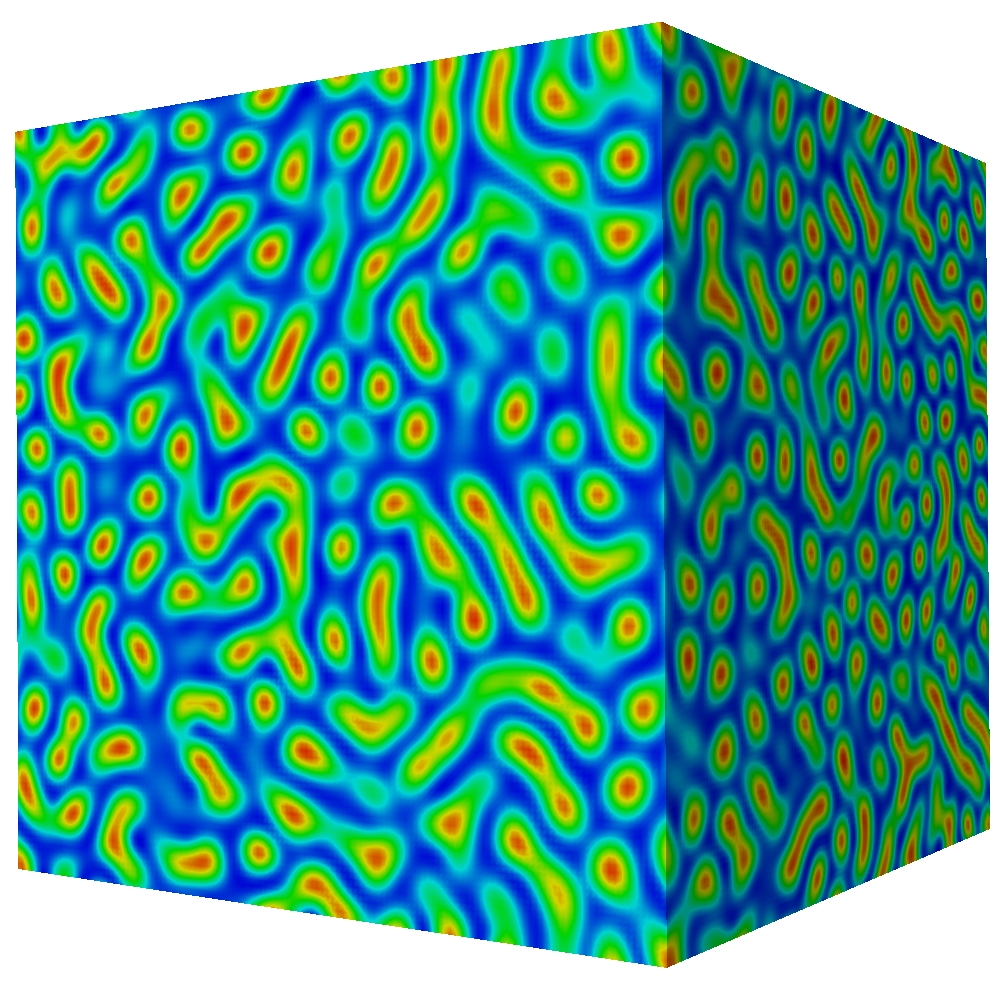}
\includegraphics[width=0.32\hsize]{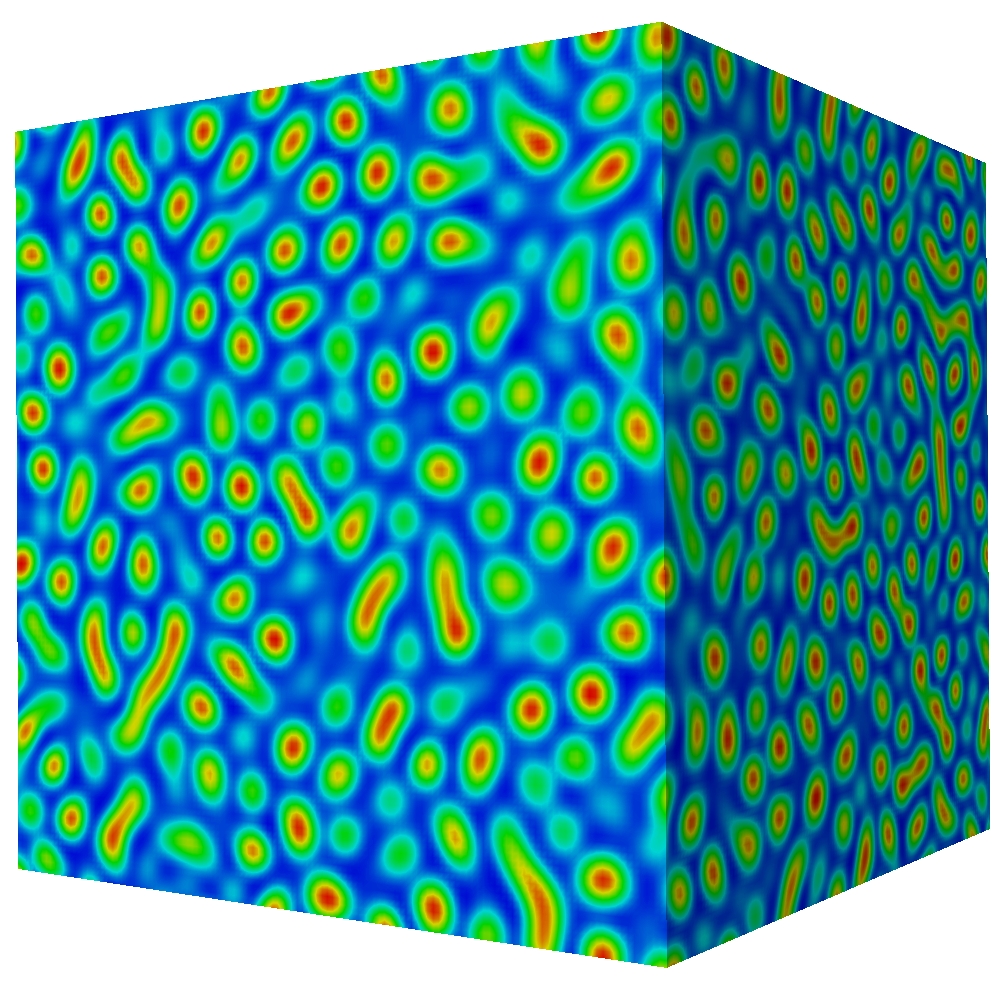}
\includegraphics[width=0.32\hsize]{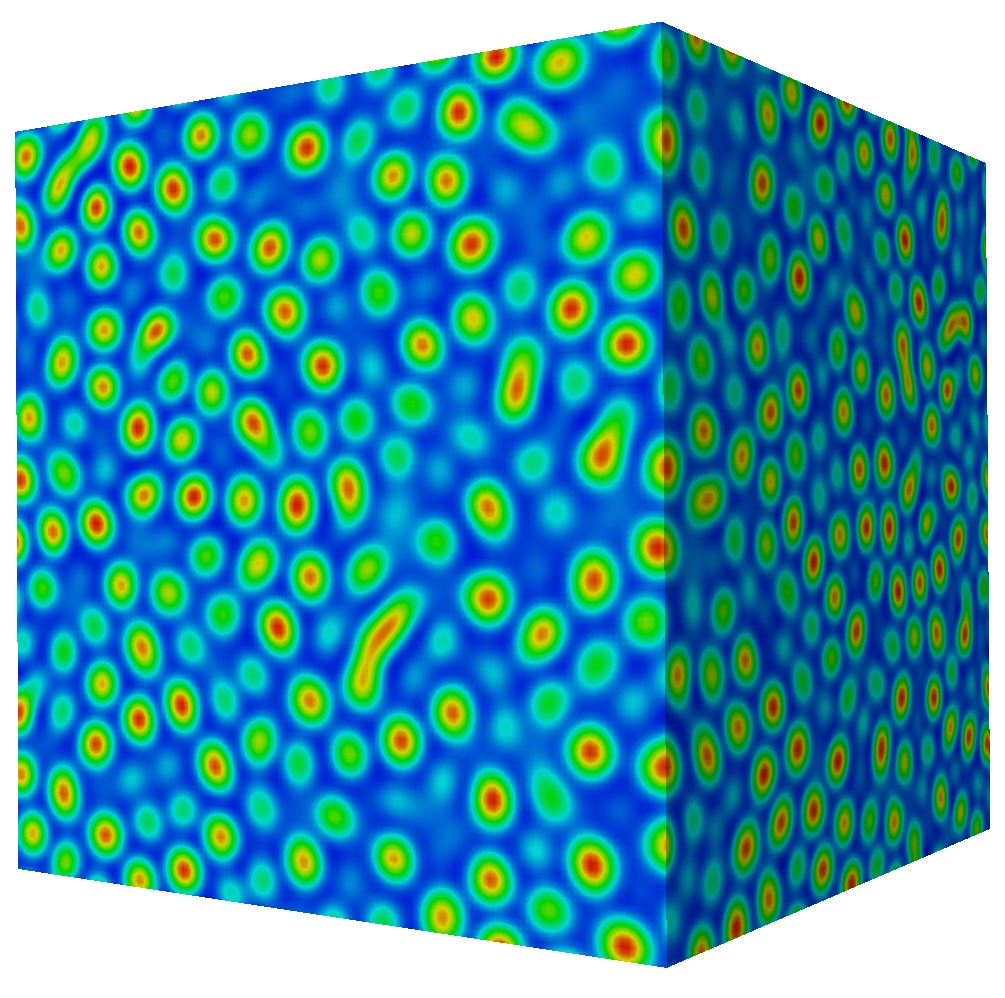}\\
\includegraphics[width=0.32\hsize]{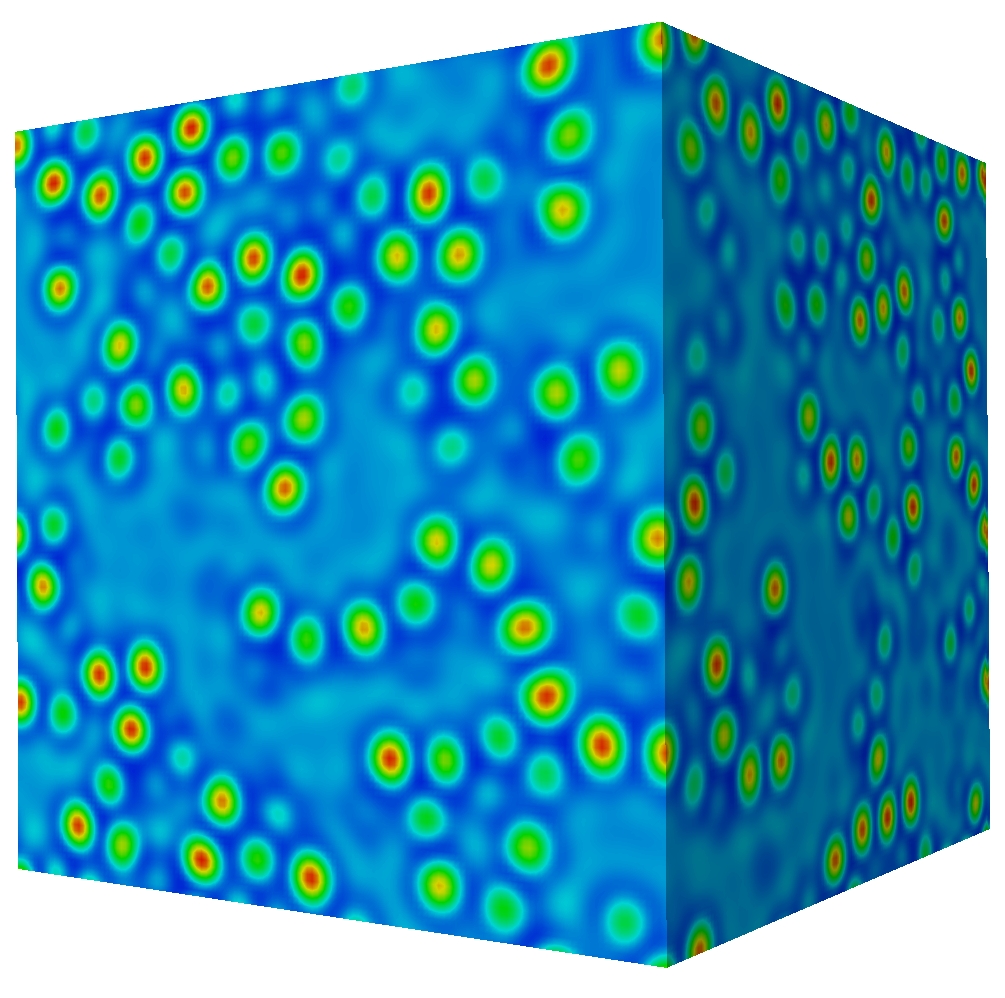}
\includegraphics[width=0.32\hsize]{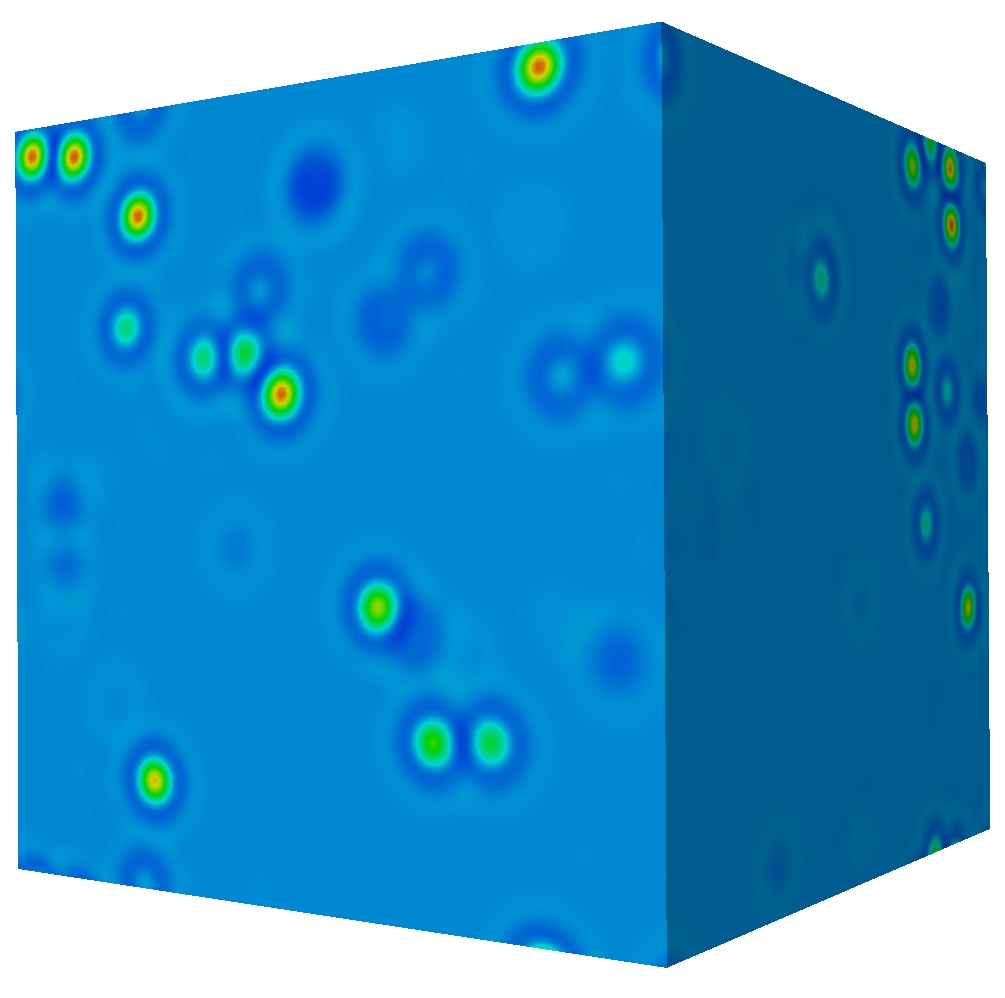}
\includegraphics[width=0.32\hsize]{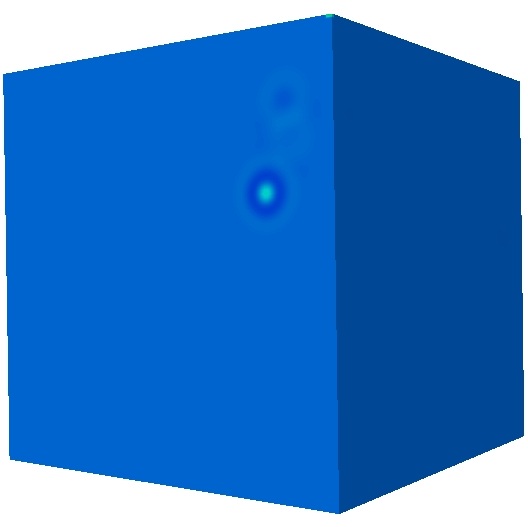}\\
\caption{(color online) Steady-state localized solutions of Eq.~(\ref{eq:csh})
in three dimensions for $r=-0.9$ and different mean order parameter values $\phi_0$: from top left to bottom right
$\phi_0=0.025, 0.125, 0.225, 0.325, 0.425, 0.525, 0.625, 0.725$ and $0.750$.
The domain size is $100\times100\times100$.}
\mylab{fig:prof-threedim-one}
\end{figure}

In Fig.\ \ref{fig:loc-2d-fam-rm09} we superpose the 2d and 3d DNS results for $r=-0.9$ on the 1d bifurcation diagrams for $r=-0.9$ and system size $L=100$ (Fig.~\ref{fig:loc-fam-rm09}). The 2d order parameter profiles are calculated on a domain with area $L^2=100^2$ and the 3d results have a system volume $L^3=100^3$. In Fig.\ \ref{fig:loc-2d-fam-rm09}(a) we display the $L^2$ norm $||\delta\phi ||$, in (b) the chemical potential $\mu$, in (c) the mean free energy $(F-F_0)/L^d$, and in (d) the mean grand potential $\omega=F/L^d-\phi_0\mu$. In each plot the connected (violet) squares correspond to results from 2d calculations such as those displayed in Figs.~\ref{fig:cSH-phasediagram-2d} and \ref{fig:prof-twodim-one} and the connected green triangles correspond to 3d results such as those displayed in Fig.~\ref{fig:prof-threedim-one}. It is from examining the figures in panels (c) and (d) that one can most easily discern the reason for the main differences between the 1d results and the 2d and 3d results: we see (particularly in the 2d results) that the chemical potential $\mu$ and the pressure $p=-\omega$ have regions where these measures are roughly flat as a function of $\phi_0$ and regions where they increase as a function of $\phi_0$. We also see that these increases in some places relate to features in the 1d results but in other places they do not have any relation to what one sees from the 1d results. This is because in 2d and 3d the system displays phases that are not seen in 1d (cf.~Figs.\ \ref{fig:cSH-phasediagram-1d} and \ref{fig:cSH-phasediagram-2d}). To understand the origin of these roughly flat portions, we recall that in the thermodynamic limit $L\to\infty$ two states are said to be at coexistence if the `temperature' $r$, the chemical potential $\mu$ and the pressure $p=-\omega$ is the same for both states. These quantities do not change in value as one takes the system across the coexistence region by increasing the average density in the system (or equivalently $\phi_0$). This is because the additional surface excess free energy terms (surface tension terms that are present because both coexisting phases are in the system) do not contribute in the thermodynamic limit. This is in turn a consequence of the fact that these surface terms scale as $L^{(d-1)}$, whereas the bulk volume terms scale as $L^d$, where $d$ is the dimensionality of the system. Thus, regions where these measures are approximately flat are in a coexistence region between two phases -- this can be confirmed for the 2d results by comparing the ranges of $\phi_0$, where the results for $\mu$ and $\omega$ are approximately flat, with the coexistence regions in Fig.\ \ref{fig:cSH-phasediagram-2d}. The observation that the 2d curves in Fig.\ \ref{fig:loc-2d-fam-rm09} are not completely flat indicates that the interfacial (surface tension) terms between the different phases in the LS state do contribute to the free energy, and is thus a finite size effect. Note that it might be possible to distribute the LS in such a way that they percolate throughout the whole system so that the contribution from the interfaces scales as $L^d$. If this is the case, the above argument does not apply.

\begin{figure}
\textbf{(a)}\includegraphics[width=0.45\hsize]{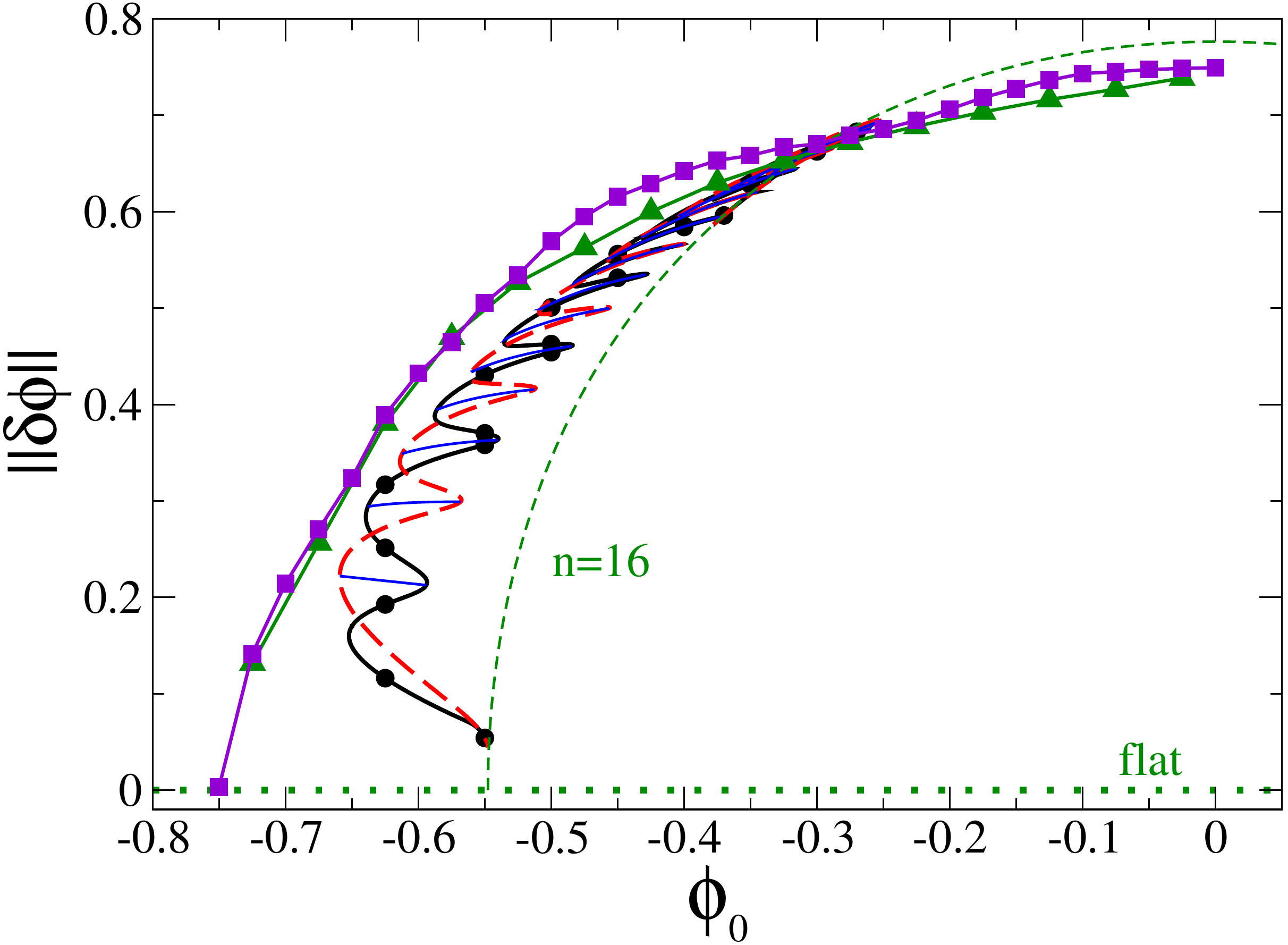}
\includegraphics[width=0.45\hsize]{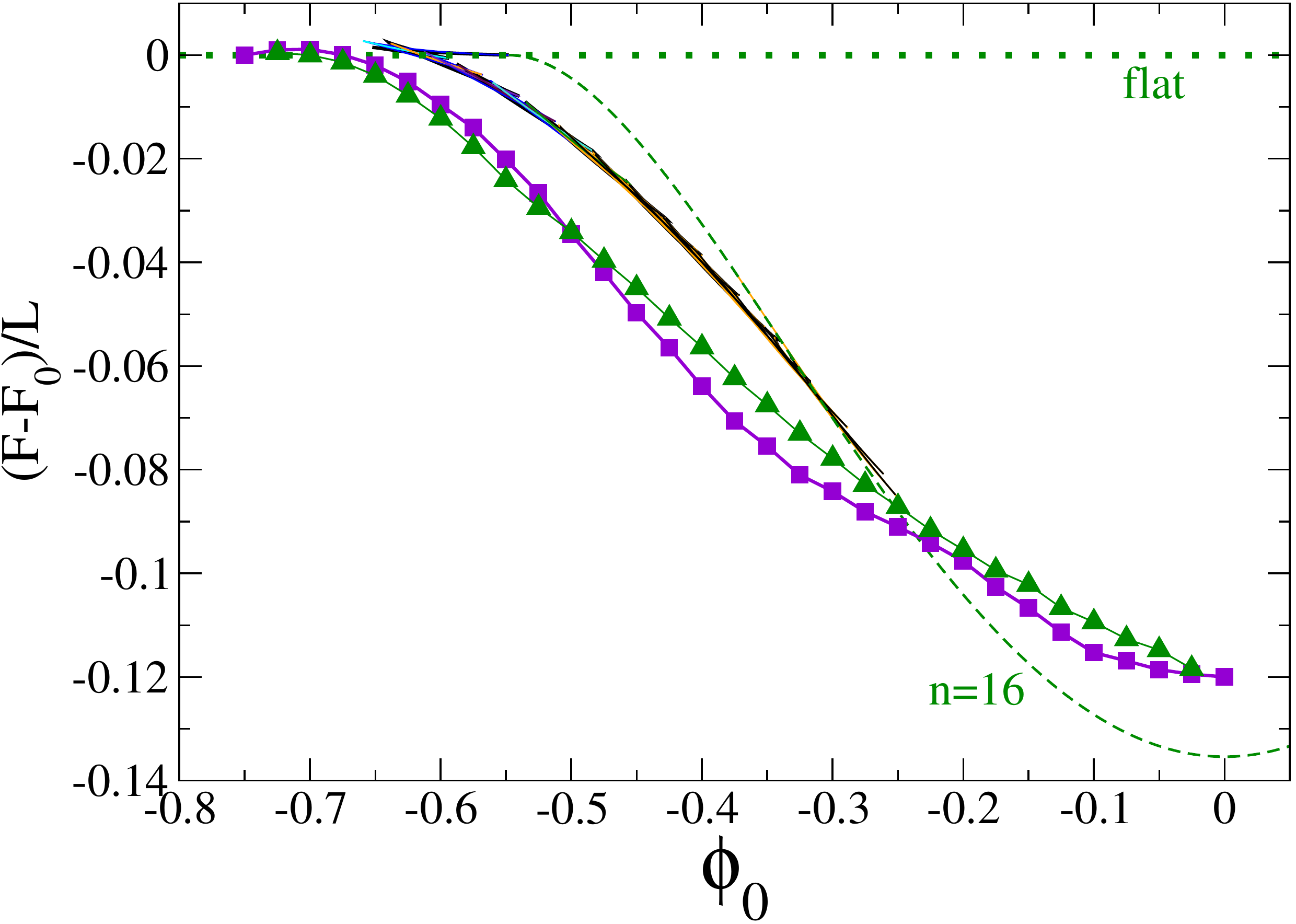}\textbf{(b)}
\textbf{(c)}\includegraphics[width=0.45\hsize]{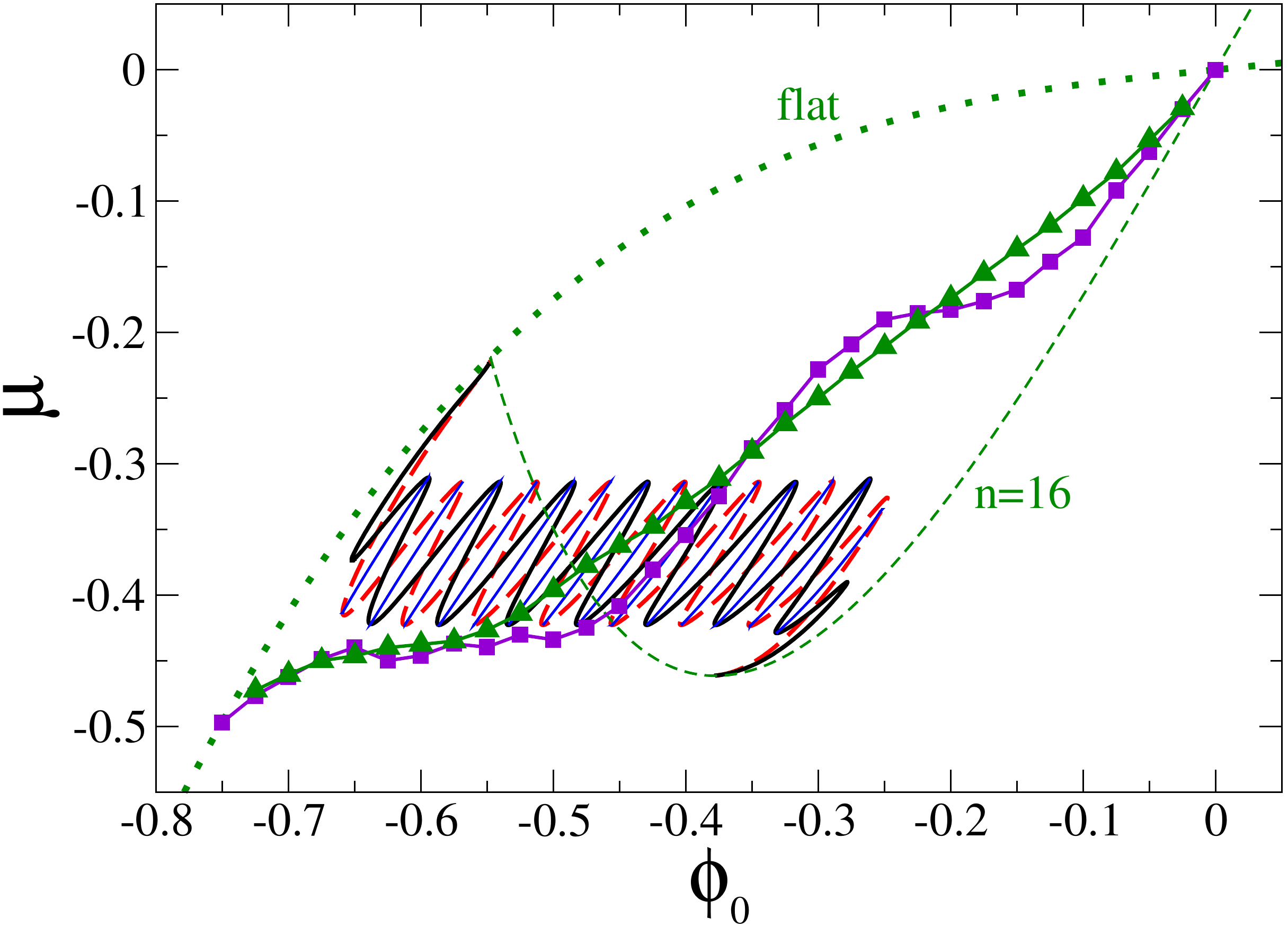}
\includegraphics[width=0.45\hsize]{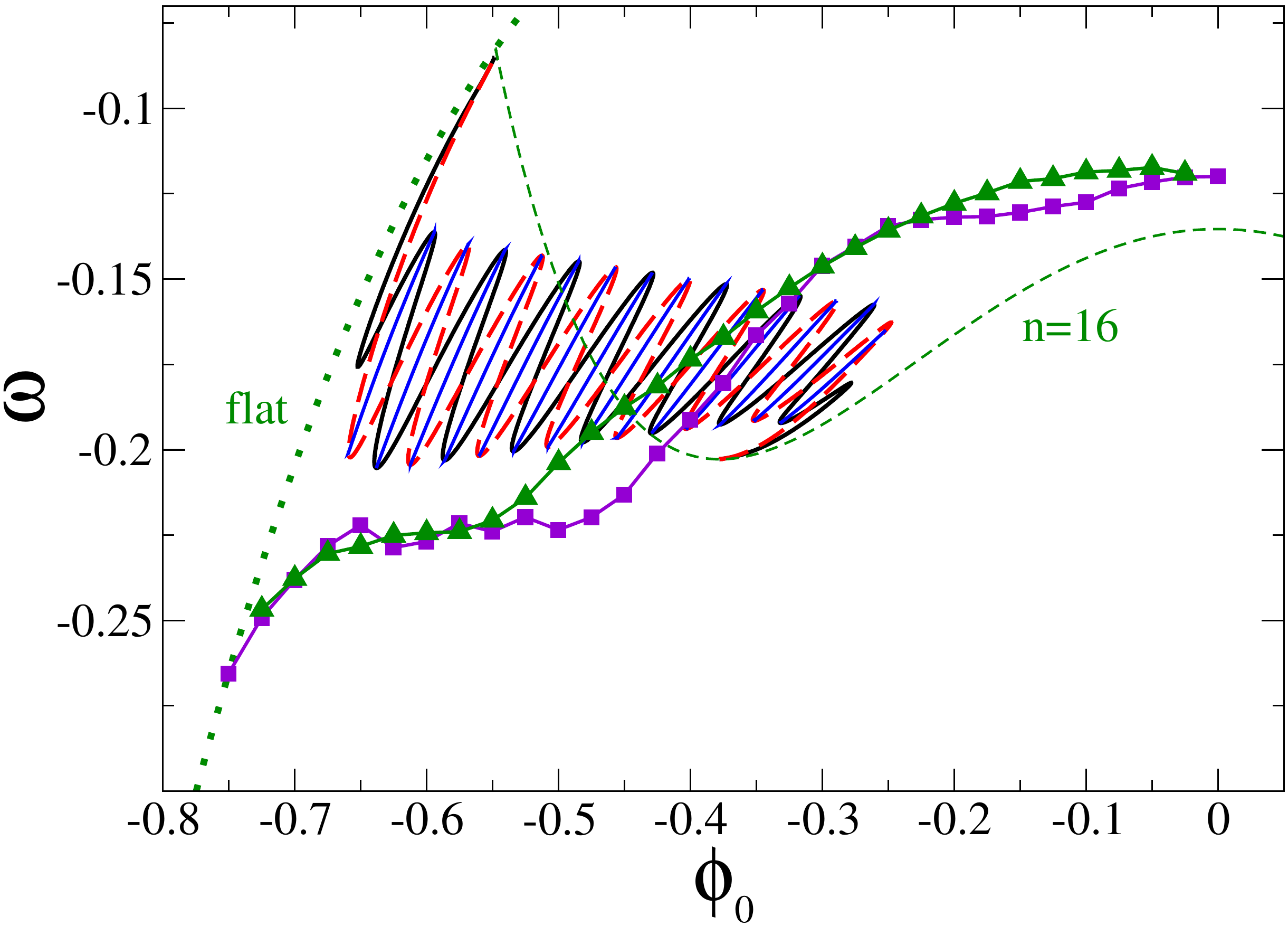}\textbf{(d)}
\caption{(color online) Characteristics of steady-state localized solutions
    of the conserved Swift-Hohenberg equation for $r=-0.9$, as a
    function of the mean order parameter $\phi_0$ on fixed one,
    two and three-dimensional domains of size $L^d$, $d=1,2,3$,
    and $L=100$.  The various solution profiles are
    characterized by their (a) $L^2$ norm $||\delta\phi||$, (b)
    chemical potential $\mu$, (c) mean free energy $(F-F_0)/L^d$, and
    (d) mean grand potential $\omega=F/L^d-\phi_0\mu$.
The connected (violet) square and (green) triangle symbols correspond
to 2d calculations (for sample profiles see
Figs.~\ref{fig:cSH-phasediagram-2d} and \ref{fig:prof-twodim-one}) and
3d calculations (for sample profiles see
Fig.~\ref{fig:prof-threedim-one}). For comparison we also show the 1d
results from Fig.~\ref{fig:loc-fam-rm09} for the periodic (i.e.,
stripe) state with $n=16$ bumps, green dashed) and the various 1d
localized states. The thick green dotted line corresponds to the
homogeneous solution $\phi(\mathbf{x})=\phi_0$.}  \mylab{fig:loc-2d-fam-rm09}
\end{figure}

\section{Discussion and conclusions}

The conserved Swift-Hohenberg equation is perhaps the simplest example of a
pattern-forming system with a conserved quantity. Models of this type arise
when modelling a number of different systems, with the PFC model being one 
particular example. Other examples include binary fluid convection between 
thermally insulating boundaries \cite{K:89}, where this equation
was first derived, convection in an imposed magnetic field (where the 
conserved quantity is the magnetic flux \cite{CM01,LBK11}) and two-dimensional 
convection in a rotating layer with stress-free boundaries (where the conserved 
quantity is the zonal velocity \cite{CM01,BBKK12}). Models of a vibrating layer
of granular material are also of this type (here the conserved quantity is the
total mass \cite{DL10}). It is perhaps remarkable that all these distinct 
systems behave very similarly. In particular, they all share the following 
features:
(i) strongly subcritical bifurcations forming localized structures so that 
the resulting LS are present outside of the bistability region between the 
homogeneous and periodic states; (ii) presence of LS even when the periodic 
branch is supercritical; (iii) organization of LS into slanted snaking; and 
(iv) the transition from slanted snaking to smooth snaking whereby the LS 
grow smoothly without specific bump-forming events (referred to as 
`nucleation' events in the pattern formation literature \footnote{In the
condensed matter literature the word `nucleation' refers to the traversing 
of a (free) energy barrier to form a new phase. In the theory of pattern 
formation, this term is used more loosely to describe the appearance or 
birth of a new structure, bump, etc, without necessarily implying that there 
is an energy barrier to be crossed.}.) 
These properties of the system can all be traced to the fact that the conserved 
quantity is necessarily redistributed when an instability takes place or a
localized structure forms. This fact makes it harder for additional LS to 
form and as a result the system has to be driven harder for this to occur, 
leading to slanted snaking. Bistability is no longer required since the
localized structures are no longer viewed as inclusions of a periodic state
within a homogeneous background or vice versa. These considerations also
explain why the bifurcation diagrams in these systems are sensitive to the
domain size, and it may be instructive, although difficult, to repeat some
of our calculations for larger domain sizes.

\begin{figure}
\includegraphics[width=0.9\hsize]{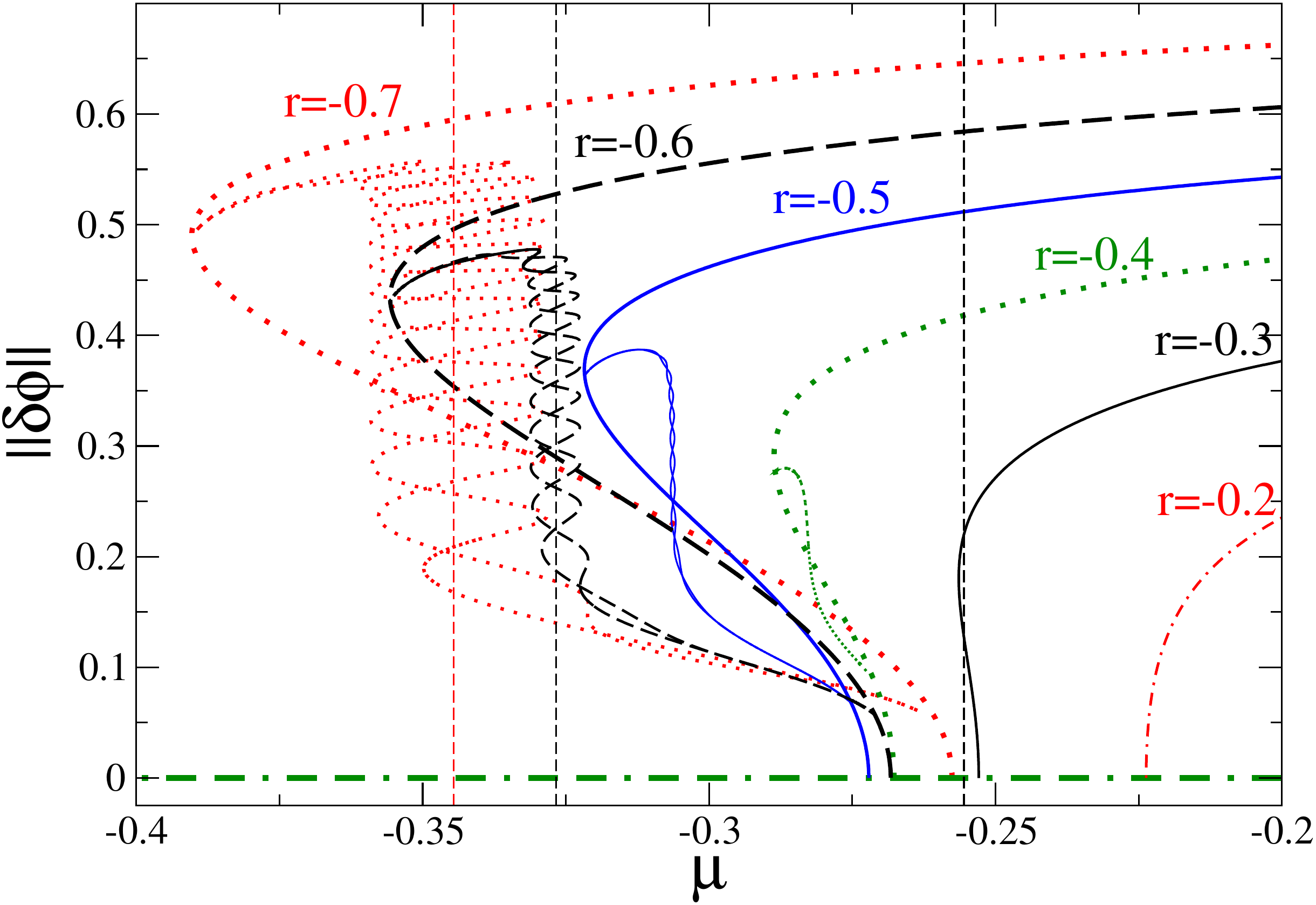}  
\caption{(color online) The $L^2$ norm $||\delta\phi ||$ for the
    homogeneous, periodic and localized steady state solutions of the
    conserved Swift-Hohenberg equation (\ref{eq:csh}) as a function of the
    chemical potential $\mu$, for a fixed domain size of $L=100$ and
    various $r$. The horizontal thick dot-dashed green 
    line corresponds to the homogeneous solution
    $\phi(x)=\phi_0$. Periodic solutions with $n=16$ peaks are labeled
    by the corresponding $r$ values, whereas the branches that
    bifurcate from the periodic solution branches represent the two
    types of symmetric localized states (LS$_\mathrm{odd}$ and
    LS$_\mathrm{even}$): $r=-0.7$ (dotted red lines), $r=-0.6$ (dashed
    black lines), $r=-0.5$ (solid blue lines), $r=-0.4$ (dotted green 
    lines), $r=-0.3$ (solid black line) and $r=-0.2$ (dashed red line). 
    The dashed vertical lines indicate the corresponding coexistence
    chemical potential values
    for $r=-0.7$, $-0.6$ and $-0.3$, respectively.}
\mylab{fig:bif-over-mu}
\end{figure}

Of particular significance is the observation that the slanted snaking
found when the LS norm $||\delta \phi ||$ is plotted as a function of
the mean order parameter $\phi_0$ for fixed $r$
(Figs.~\ref{fig:loc-fam-rm09} and \ref{fig:loc-fam-sevrm-one}(a,b)) is
`straightened' out when the same solution branches are displayed as a
function of $\mu$ (see Fig.~\ref{fig:bif-over-mu}). The snaking
is then vertically aligned and centred around the coexistence chemical
potential values (referred to as Maxwell points in Ref.\ \cite{BuKn06}) as
known from the non-conserved Swift-Hohenberg equation \cite{BuKn06}.  In
this representation the periodic states typically bifurcate subcritically 
from the homogeneous state (Fig.~\ref{fig:bif-over-mu}), in contrast to 
the supercritical transitions found when $\phi_0$ is used as the bifurcation 
parameter (Figs.~\ref{fig:loc-fam-rm09} and \ref{fig:loc-fam-sevrm-one}). 
It is also worth noting that the thermodynamic tricritical point at
$(\phi_{0b},r_b^\mathrm{max})=(\pm\sqrt{3/38},-9/38)$ discussed in
Sec.~\ref{sec:loc-states-1d} (see also the last paragraph of the Appendix)
corresponds to the transition from a
subcritical to a supercritical primary bifurcation in the
representation of Fig.~\ref{fig:bif-over-mu}, i.e., when $\mu$ is used
as the bifurcation parameter. This transition takes place between the
solid black line for $r = -0.3$ and the dashed red line for $r = -0.2$
in Fig.~\ref{fig:bif-over-mu} and implies that the linear stability
properties of \textit{identical} periodic solutions differ, depending
on whether the permitted perturbations preserve $\mu$ (and hence
permit $\phi_0$ to vary) or vice versa, and this is so for the
localized states as well.  In particular, when $\mu$ is used as the
control parameter, the differences identified in Sec.~\ref{sec:snake}
between cases (i)-(iv) where the LS branches do not snake, and cases
(v)-(vi) where they do snake disappear. In Fig.~\ref{fig:bif-over-mu}
snaking over $\mu$ is clearly visible for $r=-0.7$, $-0.6$ and $-0.5$.
There is no snaking for $r=-0.4$, most likely due to finite size
effects, or for larger $r$.

The thermodynamic reason that the snaking becomes straightened when
displayed as a function of the chemical potential $\mu$ is related to
the issues discussed at the end of Sec.~\ref{sec:2d3d}. For a system
with $(\phi_0,r)$ chosen so that it is in the coexistence region
(e.g.,~between the homogeneous and the bump states) and size $L$ large
enough to be considered to be in the thermodynamic limit $L \to
\infty$, the chemical potential does not vary when $\phi_0$ is changed
and neither does the grand potential $\omega=-p$.  This is because the
free energy contribution from interfaces between the coexisting phases
(the interfacial tension) scales with the system size~$\sim
L^{(d-1)}$, and hence is negligible compared to the bulk contributions
which scale~$\sim L^d$. As $\phi_0$ is varied so as to traverse the
coexistence region, new bumps are added or removed from the bump state
(LS). For a (thermodynamically) large system, the resulting changes in
the free energy are negligible, although this is no longer true of a
finite size system. Furthermore, the interfacial free energy
contribution between the two phases varies depending on the size of
the bumps right at the interface and the size of these bumps depends
on the value of $\phi_0$. The difference between the maximal
interfacial energy and the minimal value determines the width of the
snake. All of these contributions have leading order terms that
scale~$\sim L^{-1}$. In a related situation, the variation in the
chemical potential as the density is varied in a finite-size system
containing a fluid exhibiting gas-liquid coexistence is discussed in
Refs.~\cite{MSE06,SVB09,BBVT12}. It is instructive to compare, for
example, Fig.~3 in Ref.\ \cite{BBVT12} and
Fig.~\ref{fig:loc-2d-fam-rm09}(c) of the present paper.

We mention that the steady states of the nonconserved Swift-Hohenberg
equation studied, e.g., in Refs.~\cite{BuKn06,BBKM08},
correspond to solutions of Eq.~(\ref{eq:csh-loc-steady}) with the
nonlinearity $g_{23}$ but $\mu=0$ (see Sec.~\ref{sec:eqs}). These
always show vertically aligned snaking when the LS norm $||\delta\phi||$ 
is plotted as a function of $r$ (adapting $\phi_0$). However, when
$\mu=0$ no localized states are present with the pure cubic nonlinearity 
employed in Eq.~(\ref{eq:csh-loc-steady}). To find such states $\mu$ must
be fixed at a value sufficiently far from zero; the resulting LS then 
exhibit vertically aligned snaking when plotted as a function of $r$
\cite{BuKn06}.

The domain size we have used, $L=100$, is moderately large. It contains 
of the order of 16 wavelengths of the primary structure-forming instability.
Because the equation is simpler than the hydrodynamic equations for
which similar behavior was observed, the results we
have been able to obtain are substantially more complete, even in one
dimension, than was possible elsewhere. In particular, we have been able to compute
the rungs of asymmetric LS and to study their behavior as the system
transitions from slanted snaking to smooth snaking as $r$ increases towards
zero. The extension of our results to two and three spatial dimensions is
necessarily incomplete, although the transition to clusters of bumps 
followed by isolated bumps as the total ``mass'' decreases is not surprising.
However, the transition from a connected structure to a disconnected one
(the ``percolation'' threshold) in two and three dimensions deserves a much 
more detailed study than we have been able to provide.

In this connection we mention two experimental systems exhibiting a 
transition from a solid-like phase to a gas-like phase of individual 
spots. This is the gaseous discharge system studied by H.-G. Purwins 
and colleagues \cite{astrov01,PBA10} and the liquid crystal light valve 
experiment of S. Residori and colleagues \cite{BCR09}. In both these 
systems a crystal-like structure of spots with hexagonal coordination 
was observed to melt into a `gas' of individual spots as a parameter 
was varied. This two-dimensional process leads to states resembling 
those found here in Figs.~\ref{fig:cSH-phasediagram-2d} and
\ref{fig:prof-twodim-one} although the stripe-like structures were typically
absent. In these two systems the spots in the `gas-phase' are mobile unlike
in the cSH equation indicating absence of variational structure. However, both
systems are globally coupled, by the imposed potential difference in the 
discharge system and the feedback loop in the liquid crystal light valve
experiment, raising the possibility that the global coupling in these 
systems plays a similar role to the role played by the conserved order
parameter in the cSH equation.

We should also mention that some of the localized states observed in Figs.\
\ref{fig:prof-twodim-one} and \ref{fig:prof-threedim-one} raise some concern
about the validity of the PFC as a model for solidification and freezing -- we
refer in particular to the order parameter profiles displayed at the bottom 
right of these figures. These show that in both 2d and 3d the PFC predicts
the existence of steady states with isolated single bumps. Recall that
in the standard interpretation of the PFC, the bumps correspond to frozen
particles whilst the homogeneous state corresponds to the uniform liquid state. 
Such profiles could perhaps be a signature of the dynamical heterogeneity that 
is a feature of glassy systems, but there are problems with this interpretation
-- the glass transition is a collective phenomenon -- single particles do not 
freeze on their own whilst the remainder of the particles around them remain 
fluid! We refer readers interested in the issue of the precise interpretation 
of the order parameter in the PFC to the discussion in the final section of 
Ref.\ \cite{RATK12}. We should also point out that although these structures 
correspond to local minima of the free energy (i.e., they are stable) they do 
not correspond to the global minimum. These states occur for state points 
where the global free energy minimum corresponds to the uniform homogeneous
state.

All the results presented here have been obtained for the generic conserved 
Swift-Hohenberg equation, Eq.~(\ref{eq:DDFT_PFC}), with the energy functional 
in Eq.\ (\ref{eq:hfe}). We believe that our main results provide a qualitative 
description of a number of related models in material science that are of a 
similar structure and describe systems that may show transitions between 
homogeneous and patterned states characterized by a finite structure length. 
In particular, we refer to systems that can be described by conserved 
gradient dynamics based on an underlying energy functional that features a 
local double-well contribution, a destabilizing squared gradient term and a
stabilizing squared Laplacian term. The latter two terms may themselves 
result from a gradient expansion of an integral describing nonlocal 
interactions such as that required to reduce DDFT models to the simpler 
PFC model~\cite{vBVL09,ARTK12}.

Other systems, where the present results may shed some light, include diblock 
copolymers and epitaxial layers. The time evolution equation for 
diblock copolymers \cite{OoSh87,Paqu91,TeNi02,Glas10,Glas12} is of fourth 
order like the nonconserved SH equation but contains a global coupling term that
is related to mass conservation. The equation emerges from a nonlocal term in
the energy functional \cite{Leib80,OhKa86}.  The global coupling results 
in an evolution towards a state with a given mean value for the density
order parameter $\phi_0$ if the initial value is different from $\phi_0$
or in a conservation of mass as the system evolves if the initial value 
coincides with the imposed $\phi_0$. Although this differs from the 
formulation using a conserved Swift-Hohenberg equation, the steady versions 
of the diblock-copolymer equation and of the conserved Swift-Hohenberg 
equation are rather similar: they only differ in the position of the
nonlinearity. Up to now no systematic study of localized states
exists for the diblock-copolymer equation, although Ref.~\cite{Glas12} 
discusses their existence and gives some numerical examples
for a profile with a single bump in rather small systems (see their Fig.~5).
Since in the diblock-copolymer system the order parameter is a conserved
quantity, we would expect the snaking of localized states for fixed $\phi_0$
to be slanted similar to our Fig.~\ref{fig:loc-fam-rm09}, instead of being 
vertical, corresponding to a standard snake, where all the saddle-node
bifurcations are vertically aligned, as sketched in Fig.~6 of \cite{Glas12}.

Finally, we briefly mention a group of model equations that
are derived to describe the evolution of the surface profile of
epitaxially strained solid films including, e.g., the
self-organization of quantum dots \cite{GDV03}. The various
evolution equations that have been employed account for the 
elasticity (linear and  nonlinear isotropic elasticity, as well as 
misfit strain) of the epitaxial layer and the (isotropic or anisotropic)
wetting interaction between the surface layer and the solid beneath. 
The evolution equations we wish to highlight are of sixth order
\cite{SDV93,Savi03,GLSD04,Thie10} much like the conserved
Swift-Hohenberg equation investigated here. Other models, however,
are of fourth order only \cite{XiE02,SRF03} or contain fully nonlocal
terms (resulting in integro-differential equations)
\cite{XiE04,LGDV07}. However, even the sixth order models often
contain additional nonlinear terms in the derivatives (see, for
instance, Eq.~(5) of Ref.~\cite{GDV03}). Localized state solutions of
these equations have to our knowledge not yet been studied
systematically, although some have been obtained numerically (Fig.~5
of \cite{GDV03}). Future research should investigate how the
characteristics of the localized states analysed here for the
conserved Swift-Hohenberg equation differ from those in
specific applied systems such as diblock copolymers or epitaxial
layers.  

\section*{Appendix}

In this Appendix we determine the direction of branching of the
localized states when they bifurcate from the branch of periodic
states. When the domain is large this bifurcation occurs when the
amplitude of the periodic states is small and hence is accessible to
weakly nonlinear theory.

We begin with Eq.~(\ref{eq:csh-loc}) which may be written
\begin{equation}
\phi_t=\alpha\partial_x^2[(r+q^4)\phi+\phi^3+2q^2\partial_x^2\phi+\partial_x^4\phi].
\end{equation}
This equation has the homogeneous solution $\phi=\phi_0$. We let $\phi=\phi_0+\psi$, obtaining
\begin{equation}
\psi_t=\alpha\partial_x^2[(r+q^4+3\phi_0^2)\psi+3\phi_0\psi^2+\psi^3+2q^2\partial_x^2\psi+\partial_x^4\psi].
\end{equation}
Linearizing and looking for solutions of the form $\psi\propto\exp(\beta t+ikx)$ we obtain the dispersion relation
\begin{equation}
\beta=-\alpha k^2[r+(q^2-k^2)^2+3\phi_0^2]
\end{equation}
The condition $\beta=0$ gives the critical wavenumbers
\begin{equation}
k_c^2=q^2\pm\sqrt{-r-3\phi_0^2}
\end{equation}
Thus when $r=r_c\equiv -3\phi_0^2$ the neutral curve has maxima at both $k=0$ and $k_c=q$. When $r=-3\phi_0^2-\epsilon^2\nu$, where $\nu={\cal O}(1)$ and $\epsilon$ is a small parameter that defines how far $r$ is from $r_c$, then a band of wavenumbers near $k=q$ grows slowly with growth rate $\beta={\cal O}(\epsilon^2)$, while wavenumbers near $k=0$ decay at the same rate. There is therefore time for these two disparate wavenumbers to interact and it is this interaction that determines the direction of branching.

These considerations suggest that we perform a two-scale analysis
with a short scale $x={\cal O}(q^{-1})$ and a long scale $X=\epsilon x$, so that
$\partial_x\rightarrow\partial_x+\epsilon\partial_X$ etc. We also write
\begin{equation}
\psi=\epsilon A(X,t)e^{iqx}+ \epsilon^2B(X,t)+\epsilon^2 C(X,t)e^{2iqx}+{\rm c.c.}+{\cal O}(\epsilon^3),
\end{equation}
where the amplitudes $A$ and $C$ are complex and $B$ is real. Substituting, we obtain 
\begin{equation}
A_t=-\epsilon^2\alpha q^2(-\nu A-4q^2A_{XX}+6\phi_0AB+6\phi_0C A^*+3|A|^2A)+{\cal O}(\epsilon^3),
\end{equation}
\begin{equation}
B_t=\epsilon^2\alpha(q^4B_{XX}+6\phi_0|A|^2_{XX})+{\cal O}(\epsilon^3),\label{B}
\end{equation}
and 
\begin{equation}
C_t=-4\alpha q^2(9q^4C+3\phi_0A^2)+{\cal O}(\epsilon).
\end{equation}
The last equation implies that the mode $C$ decays on an ${\cal O}(1)$ time scale to its asymptotic value, $C=-\phi_0 A^2/3q^4+{\cal O}(\epsilon)$. The resulting equations may be written in the form
\begin{equation}
A_{t}=\nu A+4A_{XX}-\xi A\theta_X-3\biggl(1-\frac{\xi^2}{54}\biggr)|A|^2A+{\cal O}(\epsilon),\label{A}
\end{equation}
\begin{equation}
\theta_{t}=\theta_{XX}+\xi |A|^2_{X}+{\cal O}(\epsilon)\label{theta}
\end{equation}
using the substitution $B=\theta_X$, and integrating Eq.~(\ref{B}) once to obtain an equation for $\theta$. In writing these equations we have absorbed $q$ into the length scale $X$ and $\epsilon^2\alpha q^2$ into the time scale $t$, and introduced the parameter $\xi\equiv 6\phi_0/q^2<0$. The resulting equation are equivalent to the equations studied by Matthews and Cox \cite{MaCo00}.

Equations (\ref{A})--(\ref{theta}) provide a complete description of the small amplitude behavior of Eq. (\ref{eq:csh-loc}). The equations inherit a gradient structure from Eq. (\ref{eq:csh-loc}),
\begin{equation}
A_t=-\frac{\delta F}{\delta A^*},\qquad \theta_t=-\frac{\delta F}{\delta \theta},
\end{equation}
where
\begin{equation}
F[A,A^*,\theta]=\int_D\biggl\{-\nu|A|^2+4|A_X|^2+\frac{1}{2}\theta_X^2+\xi|A|^2\theta_X+\frac{3}{2}\biggl(1-\frac{\xi^2}{54}\biggr)|A|^4\biggr\}\,dX.
\end{equation}
We may write this energy in the form
\begin{equation}
F[A,A^*,\theta]=\int_D\biggl\{-\nu|A|^2+4|A_X|^2+\frac{1}{2}\bigg[(\theta_X+\xi|A|^2)^2+3\biggl(1-\frac{19\xi^2}{54}\biggr)|A|^4\biggr]\biggr\}\,dX,\label{energy}
\end{equation}
implying that the free energy $F[A,A^*,\theta]$ is not bounded from below once $\xi^2>54/19$ (equivalently $\phi_0^2>3q^4/38$). This is a reflection of the presence of subcritical branches. Indeed, the steady states of Eqs.~(\ref{A})--(\ref{theta}) correspond to critical points of this energy and satisfy the {\it nonlocal} equation
\begin{equation}
4A_{XX}+(\nu-\langle|A|^2\rangle)A-3\biggl(1-\frac{19\xi^2}{54}\biggr)|A|^2A=0,\label{nonlocal}
\end{equation}
where $\langle(\cdots)\rangle\equiv L^{-1}\int_0^L(\cdots)\,dx$ and
$L$ is the domain length. This equation demonstrates (i) that the
primary bifurcation is subcritical when $\xi^2>54/19$ in agreement
with Eq.~(\ref{energy}), and (ii) that as $|A|$ increases the value of $\nu$ has to be raised in order to maintain the same value of the effective bifurcation parameter ${\nu}_{\rm eff}\equiv\nu-\langle|A|^2\rangle$. This is the basic reason behind the slanted structure in Fig.~\ref{fig:loc-fam-rm09}(a). Equations of this type arise in numerous applications \cite{hall,Elmer88a,Elmer88b,BBKK12} and their properties have been studied in several papers \cite{Elmer88a,norbury02,vega05,norbury07}. We mention, in particular, that unmodulated wavetrains bifurcate supercritically when $\xi^2<54$, a condition that differs from the corresponding condition $\xi^2<54/19$ for spatially modulated wavetrains.

Equations (\ref{A})--(\ref{theta}) possess the solution $(A,\theta)=(A_0,0)$, where $|A_0|^2=\nu/[3(1-\xi^2/54)]$, corresponding to a spatially uniform wavetrain. This state bifurcates in the positive $\nu$ direction wherever $\xi^2<54$, or equivalently $\phi_0^2<3q^4/2$. In the following we are interested in the modulational instability of this state. We suppose that this instability takes place at $\nu=\nu_c$ and write $\nu=\nu_c+\delta^2{\tilde \nu}$, where $\delta\ll\epsilon$ is a new small parameter and ${\tilde \nu}={\cal O}(1)$. In addition, we write $A=A_0(1+{\tilde A})$, $\theta_{X}={\tilde V}$. Since the imaginary part of ${\tilde A}$ decays to zero we take ${\tilde A}$ to be real and write
\begin{equation}
{\tilde A}=\delta{\tilde A}_1+\delta^2{\tilde A}_2+\delta^3{\tilde A}_3+\dots,\quad {\tilde V}=\delta{\tilde V}_1+\delta^2{\tilde V}_2+\delta^3{\tilde V}_3+\dots.
\end{equation}
Finally, since the localized states created at $\nu_c$ are stationary, we set the time derivatives to zero and integrate Eq.~(\ref{theta}) once obtaining
\begin{equation}
{\tilde V}+\xi |A|^2+C(\delta)=0\label{V},
\end{equation}
where $C=C_0+\delta C_1+\delta^2C_2+\delta^3C_3+\dots$ and the $C_j$ are determined by the requirement that the average of $V$ over the domain vanishes.

Substitution of the above expressions yields a sequence of ordinary differential equations which we solve subject to periodic boundary conditions. At ${\cal O}(1)$ we obtain $C_0=-\xi|A_c|^2$, where $A_c$ denotes the value of $A_0$ at $\nu=\nu_c$. At ${\cal O}(\delta)$ we obtain the linear problem
\begin{equation}
-2\nu_c {\tilde A}_1+4{\tilde A}_{1XX}-\xi {\tilde V}_1=0,\qquad {\tilde V}_1+2\xi|A_c|^2{\tilde A}_1=0, \qquad C_1=0,
\end{equation}
and conclude that ${\tilde A}_1=A_{11}\cos\ell X$, ${\tilde V}_1=V_{11}\cos\ell X$, provided that $\nu_c=-2\ell^2[(1-\xi^2/54)/(1-19\xi^2/54)]$. Note that this quantity is positive when $54>\xi^2>54/19$ (equivalently $3q^4/2>\phi_0^2>3q^4/38$) and that it vanishes in the limit $\ell\rightarrow0$, i.e., for infinite modulation length scale.

At ${\cal O}(\delta^2)$ we obtain  
\begin{equation}
-2\nu_c {\tilde A}_2+4{\tilde A}_{2XX}-\xi {\tilde V}_2=\xi {\tilde A}_1{\tilde V}_1+3\nu_c {\tilde A}_1^2,\qquad {\tilde V}_2+\xi|A_c|^2(2{\tilde A}_2+{\tilde A}_1^2)+C_2=0.
\end{equation}
Thus ${\tilde A}_2=A_{20}+A_{22}\cos 2\ell X$,  ${\tilde V}_2=V_{22}\cos 2\ell X$, where
\begin{equation}
A_{20}=-\frac{3}{4}\biggl[\frac{1-(13\xi^2/54)}{1-(\xi^2/54)}\biggl]A_{11}^2,\qquad
A_{22}=\frac{1}{4}A_{11}^2,\qquad
V_{22}=-\xi|A_c|^2A_{11}^2,
\end{equation}
together with
\begin{equation}
C_2=-\frac{1}{2}\xi|A_c|^2A_{11}^2-2\xi|A_c|^2A_{20}.
\end{equation}

Finally, at ${\cal O}(\delta^3)$ we obtain  
\begin{equation}
-2\nu_c {\tilde A}_3+4{\tilde A}_{3XX}-\xi {\tilde V}_3=2{\tilde\nu}{\tilde A}_1+\xi ({\tilde A}_1{\tilde V}_2+{\tilde A}_2{\tilde V}_1)+\nu_c (6{\tilde A}_1{\tilde A}_2+{\tilde A}_1^3),
\end{equation}
and 
\begin{equation}
{\tilde V}_3+\xi|A_c|^2(2{\tilde A}_3+2{\tilde A}_1{\tilde A}_2)+C_3=-\frac{2\xi}{3(1-(\xi^2/54))}{\tilde\nu}{\tilde A}_1.
\end{equation}
The direction of branching of solutions with ${\tilde A}\ne0$ follows from the solvability condition for ${\tilde A}_3$, i.e., the requirement that the inhomogeneous terms contain no terms proportional to $\cos\ell X$. We obtain
\begin{equation}
{\tilde\nu}+3\ell^2{\tilde A}_{11}^2\biggl[\frac{1-(13\xi^2/54)}{1-(19\xi^2/54)}\biggr]=0.\label{direction1}
\end{equation}
It follows that the localized states bifurcate in the positive $\nu$
direction (lower temperature) when $54/19<\xi^2<54/13$ and in the negative
$\nu$ direction when $54/13<\xi^2<54$. The former requirement is equivalent
to  $3q^4/38<\phi_0^2<3q^4/26$, the latter to $3q^4/26<\phi_0^2<3q^4/2$. These 
results agree with the corresponding results for an equation similar to 
Eq.~(\ref{nonlocal}) obtained by Elmer \cite{Elmer88a} (see also \cite{MaCo00,BBKK12}).

In the above calculation we
have fixed the parameter $\xi$ (equivalently $\phi_0$) and treated $r$
(equivalently $\nu$) as the bifurcation parameter. However, we can
equally well do the opposite. Since the calculation is only valid in
the neighborhood of the primary bifurcation at $r=-3\phi_0^2$, i.e.,
in the neighborhood of $\xi=-\sqrt{-12r}/q^2$, the destabilizing
perturbation analogous to ${\tilde\nu}>0$ is now replaced by
${\tilde\xi}>0$ (cf.~Fig.~\ref{fig:cSH-phasediagram-1d}), implying that
the direction of branching of localized states changes from subcritical 
to supercritical as $\xi^2$ decreases through $\xi^2=54/13$ (equivalently
$\phi_0^2$ decreases through $\phi_0^2=3q^4/26$ or $r$ increases through
$r=-9q^4/26\approx-0.35q^4$), in good agreement with the numerically
obtained value $r\approx -0.41$ (for $q=1$) that is indicated in
Fig.~\ref{fig:prof-loc-folds-zoom} by the horizontal line separating
regions (iii) and (iv).

We next consider the corresponding results in the case when the chemical 
potential $\mu$ is fixed. In this case the constants $C_j$ vanish, and
the ${\cal O}(\delta^2)$ problem therefore has solutions of the form
${\tilde A}_2=A_{20}+A_{22}\cos 2\ell X$, ${\tilde V}_2=V_{20}+V_{22}\cos 2\ell X$,
where
\begin{equation}
A_{20}=-\frac{3}{4}A_{11}^2,\qquad V_{20}=\xi |A_c|^2 A_{11}^2,\qquad
A_{22}=\frac{1}{4}A_{11}^2,\qquad V_{22}=-\xi|A_c|^2A_{11}^2.
\end{equation}
The nonzero $V_{20}$ contributes additional terms to the solvability
condition at ${\cal O}(\delta^3)$. The result corresponding to 
(\ref{direction1}) is 
\begin{equation}
{\tilde\nu}+3\ell^2{\tilde A}_{11}^2\biggl[\frac{1-(\xi^2/54)}{1-(19\xi^2/54)}\biggr]=0.\label{direction2}
\end{equation}
This implies that the secondary LS branches are subcritical whenever the periodic branch is supercritical ($\xi^2<54$) and the secondary bifurcation is present ($\xi^2>54/19$). These results are consistent with Fig.~\ref{fig:bif-over-mu} and moreover predict that the LS states are absent for $\xi^2<54/19$, i.e., for $\phi_0>-\sqrt{3/38}q^2$ or $r>-9q^4/38$ (cf. Fig.~\ref{fig:bif-over-mu}). This point is, of course, the thermodynamic tricritical point discussed in Sec.\ \ref{sec:loc-states-1d}.

{\bf Acknowledgement}. The authors wish to thank the EU for financial
support under grant MRTN-CT-2004-005728 (MULTIFLOW) and the National 
Science Foundation for support under grant DMS-1211953.

\end{document}